\documentclass[prd,preprintnumbers,aps,preprint,amsmath,amssymb,superscriptaddress,floatfix,nofootinbib,unsortedaddress]{revtex4}
\pdfoutput=1

\usepackage{amssymb, bm, amsmath, graphicx, physymb}
\usepackage[colorlinks=true,linktocpage=true,linkcolor=blue,citecolor=blue]{hyperref}
\usepackage[usenames,dvipsnames]{color}
\usepackage{slashed}
\usepackage[utf8]{inputenc}
\usepackage[T1]{fontenc}
\usepackage{enumerate}
\usepackage{amsfonts}
\usepackage{mathtools}
\usepackage{latexsym}
\usepackage{hyperref}
\usepackage{epstopdf} 
\usepackage{bbm}
\usepackage{soul}
\usepackage[normalem]{ulem}
\usepackage{braket}
\usepackage{nccmath}
\usepackage{graphicx}

\usepackage[e]{esvect}

\def\cP{{\cal P}}

\def\cN{{\cal N}}

\newcommand{\secn}[1]{Section~1}
\newcommand{\appn}[1]{Appendix~1}

\long\def\comment#1{ }

\def\and{\quad\text{and}\quad}

\def\q{{\boldsymbol q}}
\def\0{{\boldsymbol 0}}
\def\1{{\boldsymbol 1}}
\def\p{{\boldsymbol p}}
\def\l{{\boldsymbol l}}

\def\x{{\boldsymbol x}}

\def\r{{\boldsymbol r}}

\def\b{{\boldsymbol b}}
\def\0{{\boldsymbol 0}}

\def\rn{{\boldsymbol r}}
\def\un{{\boldsymbol u}}

\newcommand{\tr}{\mathrm{tr}}

\newcommand{\N}{\mathcal{N}}

\renewcommand\a{\alpha}
\renewcommand\b{\beta}
\renewcommand\d{\delta}
\renewcommand\l{\lambda}
\renewcommand\r{\rho}

\renewcommand\t{\tau}

\newcommand\e{\epsilon}
\newcommand\g{\gamma}
\newcommand\z{\zeta}
\newcommand\m{\mu}

\newcommand\s{\sigma}

\newcommand{\tvec}{\boldsymbol}
\renewcommand{\vec}[1]{\vv{#1}}

\renewcommand{\part}{{\rm part}}

\newcommand{\be}{\begin{equation}}
\newcommand{\ee}{\end{equation}}
\newcommand{\bes}{\begin{subequations}}
\newcommand{\ees}{\end{subequations}}
\newcommand{\bea}{\begin{eqnarray}}
\newcommand{\eea}{\end{eqnarray}}

\newcommand{\pa}{\partial}

\renewcommand{\tr}{\textrm{tr}}
\newcommand{\nn}{\nonumber \\}
\newcommand{\na}{\nabla}

\begin{document}

\title{Jet broadening in dense inhomogeneous matter}

\author{Jo\~{a}o Barata}
\email[Email: ]{joaolourenco.henriques@usc.es}
\affiliation{Instituto Galego de F{\'{i}}sica de Altas Enerx{\'{i}}as,  Universidade de Santiago de Compostela, Santiago de Compostela 15782,  Spain}
\author{Andrey V. Sadofyev}
\email[Email: ]{andrey.sadofyev@usc.es}
\affiliation{Instituto Galego de F{\'{i}}sica de Altas Enerx{\'{i}}as,  Universidade de Santiago de Compostela, Santiago de Compostela 15782,  Spain}
\affiliation{Institute for Theoretical and Experimental Physics, NRC Kurchatov Institute, Moscow 117218, Russia}
\author{Carlos A. Salgado}
\email[Email: ]{carlos.salgado@usc.es}
\affiliation{Instituto Galego de F{\'{i}}sica de Altas Enerx{\'{i}}as,  Universidade de Santiago de Compostela, Santiago de Compostela 15782,  Spain}

\begin{abstract}
In this work, we study the jet momentum broadening in an inhomogeneous dense QCD medium. The transverse profile of this nuclear matter is described within a gradient expansion, and we focus on the leading gradient contributions. The leading parton is allowed to interact multiple times with the background through the soft gluon exchanges. We derive the associated final particle distribution using both the GLV opacity series and the BDMPS-Z formalism. We further discuss 
the modified factorization of the broadening process and the initial distribution of partons produced in a hard scattering, as well as
its consequences for phenomenological applications in the context of heavy-ion collisions and deep inelastic scattering. Finally, we present the broadening probability (describing the final state effects) in several limiting regimes, and give its numerical estimates for phenomenologically motivated sets of parameters.

\end{abstract}

\maketitle
\tableofcontents

\section{Introduction}

Jets are collimated sprays of particles, produced by hadronization and branching of an energetic quark or gluon (parton), which are often found in the final state of experiments on ultrarelativistic particle collisions. If before the hadronization stage the parton cascade develops in the presence of an underlying medium, produced in the same collision, the jet substructure gets modified due to the interactions with the background. The simplest manifestation of this process is the suppression of jet energy by matter, commonly referred to as jet quenching, which has attracted significant attention in the literature~\cite{Bjorken1982EnergyLO}, for a more recent review see \cite{Mehtar-Tani:2013pia,Qin:2015srf,Blaizot:2015lma,Sievert:2018imd}. Due to their high sensitivity to the spacetime structure of the medium, jets provide a promising tomographic tool to study the real-time evolution of nuclear matter both in heavy-ion collision (HIC) and deep inelastic scattering (DIS) experiments, see e.g. \cite{Vitev:2002pf, JET:2013cls, Betz:2014cza, Apolinario:2017sob, Li:2020zbk, Li:2020rqj, Arratia:2020nxw, He:2020iow, Apolinario:2020uvt, Sadofyev:2021ohn, Du:2021pqa, Antiporda:2021hpk} and references therein. 

The jet-medium interaction can be successfully described within perturbative QCD (pQCD) supplemented with a medium model, which is usually based on a collection of medium-induced stochastic color fields 
\cite{Gyulassy:1993hr, Baier:1996sk, Zakharov:1996fv, Baier:1996kr, Zakharov:1997uu, Baier:1998kq, Gyulassy:1999zd, Gyulassy:2000fs, Gyulassy:2000er, Gyulassy:2002yv, Arnold:2002ja, Wang:2001ifa, Zhang:2003wk, Djordjevic:2003zk, Mehtar-Tani:2006vpj}. In this picture, an energetic parton interacts with the matter through multiple $t-$channel gluon exchanges. Although such interactions lead to a negligible energy depletion of the hard parton, they induce gluon bremsstrahlung resulting in an energy loss. Description of these processes within pQCD is in general a complex problem, and some simplifying assumptions are usually needed. However, the commonly considered eikonal or static source approximations lead to a decoupling of the medium evolution from the jet energy loss and substructure modification observables, for a discussion see \cite{Sadofyev:2021ohn}. Thus, to extract the properties of the underlying medium evolution and information about its structure from jet observables one needs the associated theoretical framework to be extended beyond the simplest physical regimes.

Earlier attempts to include the effects of the medium flow into the jet energy loss calculations go back to \cite{Gyulassy:2000gk, Gyulassy:2001kr,Baier:1998yf}, where the medium dilution was considered, and to \cite{Armesto:2004pt, Armesto:2004vz}, where the flow was described within a phenomenologically motivated model with an additional momentum transfer. Transverse flow effects could also be partially accounted for from purely kinematic arguments, see e.g. \cite{Baier:2006pt, Liu:2006he, Renk:2006sx}. Only more recently, the medium evolution effects on the jet-medium interactions were formally included into the Gyulassy-Levai-Vitev (GLV) opacity expansion framework \cite{Sadofyev:2021ohn}. It was shown that the medium flow and variation of its properties in the transverse directions can be treated within the same medium model of stochastic color fields induced by in-medium sources, if the sources are allowed to move and the medium properties encoded in their potentials and density are changing from point to point. In the developed formalism, the changes in the local properties of the matter are described within a gradient expansion analogous to the one used in hydrodynamics, commonly applied to describe the evolution of the quark-gluon plasma (QGP) produced in HIC. In this way, \cite{Sadofyev:2021ohn} extends the idea to describe interactions of a probe with a hydrodynamically evolving matter within the same gradient expansion introduced in holographic models for strongly interacting plasmas, see e.g. \cite{Lekaveckas:2013lha, Rajagopal:2015roa, Sadofyev:2015hxa, Li:2016bbh, Reiten:2019fta, Arefeva:2020jvo}. 

In the current manuscript, we further develop the jet-tomography toolkit, and include the effects of the leading hydrodynamic gradients at all orders in the opacity expansion. The main result of this paper is the modification of the single parton transverse momentum distribution due to the evolution in a medium of finite longitudinal extension $L$. In the limit of a longitudinally uniform matter\footnote{An extension of \eqref{eq:global} to an arbitrarily $z$-dependence of the medium is given in \eqref{eq:non-static-result}.} it is given by \eqref{eq:compare_to_BDMPS} or \eqref{dNBDMPSZ}, and reads
\begin{align}\label{eq:global}
(2\pi)^2\frac{d\N}{d^2\tvec{p}dE} & = \int d^2\x \, e^{-i\p\cdot \x} e^{-\mathcal{V}(\tvec{x}) L}\Bigg\{1-i\frac{\mathcal{V}(\tvec{x})L^3}{6E}\Bigg(\frac{\mathcal{V}'(\tvec{x})}{\mathcal{V}(\tvec{x})}\tvec\na\mu^2+\frac{1}{\r}\tvec\na\r\Bigg)\cdot\tvec\na\mathcal{V}(\tvec{x}) \notag\\
 &+i\frac{\mathcal{V}(\tvec{x})L^2}{2E}\left(\frac{\mathcal{V}'(\tvec{x})}{\mathcal{V}(\tvec{x})}\tvec\na\mu^2+\frac{1}{\r}\tvec\na\r\right)\cdot\tvec\na\Bigg\} \frac{d\N^{(0)}}{d^2\tvec{x}dE}\,,
\end{align}
where the potential $\mathcal{V}(\tvec{x})$ describes the interaction of the parton with the matter, and $\mathcal{V}'(\tvec{x})\equiv\frac{\pa}{\pa\mu^2}\mathcal{V}(\tvec{x})$. This potential is fixed by the medium model and can be related to an effective dipole cross-section. Here, we model the matter with a background field produced by color sources with number density $\r$ and screened at distances of order $\frac{1}{\m}$. The formula \eqref{eq:global} describes the momentum distribution for a final state parton, after it has been produced from an initial hard process (with large energy $E$), described by $\frac{d\cN^{0}}{d^2\tvec{p}dE}$, and propagated through a static longitudinally uniform medium with finite transverse gradients 
$\tvec \na \r$ and $\tvec \na \mu^2$ of the medium parameters. The primary effect of the gradient terms is to generate a non-trivial angular dependence in the resulting parton distribution.

The present manuscript details the derivation of \eqref{eq:global} in two commonly-employed jet quenching formalisms and discusses its properties. In Section~\ref{sec:OE}, following the previous work done in~\cite{Sadofyev:2021ohn}, we provide a derivation of \eqref{eq:global}  in the GLV framework, performing a resummation of the associated opacity series. In Section~\ref{ref:BDMPS} we show how the same result can be obtained within the amplitude-level resummed framework introduced by Baier, Dokshitzer, Mueller, Peigné, Schiff, and Zakharov (BDMPS-Z)~\cite{Baier:1996kr,Zakharov:1996fv}. Finally, in Section~\ref{ref:pheno} we explore the properties of \eqref{eq:global} in a manner suitable for direct future applications\footnote{For a reader interested in immediate phenomenological applications and not in the technical derivation, \eqref{eq:global} (or its generalization \eqref{eq:non-static-result}) and Section~\ref{ref:pheno} contain all the necessary information.} in commonly used jet quenching models~\cite{Caucal:2019uvr,Casalderrey-Solana:2014bpa,Sievert:2019cwq,Putschke:2019yrg,He:2015pra}. We summarize our findings and discuss future avenues of research in Section~\ref{ref:conclusion}. Some additional technical details are included into two appendices.

\section{The GLV Opacity Series}\label{sec:OE}

In this section, we derive the gradient corrections to the jet broadening order-by-order in opacity expansion, and resum the obtained series. As in \cite{Sadofyev:2021ohn}, we will focus on the spatial gradients of the source density and Debye mass at zero medium velocity, using scalar QCD\footnote{At eikonal accuracy the spin flips can be ignored, justifying the use of scalar QCD, see e.g. \cite{Sadofyev:2021ohn}.} to describe the underlying theory. Then, the medium-induced color field is static and reads
\begin{align}
\label{e:potl1}
    g A_\mathrm{ext}^{\l a} (q) = (2\pi)\,g^{\l 0}\,\sum_j e^{-i\left(\tvec{q}\cdot \tvec{x}_j+q_zz_j\right)}\, t_j^a\,v_j(q)
    \: \delta\left(q^0\right) \,,
\end{align}
where we use bold font for vectors in the transverse 2D space, $v_j(q)$ is a model-dependent medium potential of an individual source numerated with $j$, while $t_j^a$ and $(\tvec{x}_j,\,z_j)$ are its color generator and spatial position. Notice that the particular form of the field is derived under an assumption of large source mass. In what follows, we will consider the Gyulassy-Wang (GW) model \cite{Gyulassy:1993hr} for the potential
\begin{align}
\label{e:potl5}
    v_j (q) \equiv \frac{- g^2}{
    -q_0^2+q_\perp^2 + q_z^2 + \mu_j^2 - i \epsilon }\,,
\end{align}
where $q_\perp\equiv|\tvec{q}|$ and $\m_j$ is the Debye mass in the HIC context defined by the local medium properties around the $i$th source.

With this model for the medium, we can turn to the details of the jet-medium interaction. We start with an initial parton distribution 
\begin{align} \label{e:N0}
    E \, \frac{d\N^{(0)}}{d^2 \tvec{p} dp_z} \equiv \frac{1}{2(2\pi)^3} \left| J(p) \right|^2     \,,
\end{align}
produced by a hard-scattering event. To form a jet the leading parton has to be highly energetic $E\simeq p_z$, and we will generally work in the eikonal limit, assuming that any transverse momentum or in-medium characteristic energy scale $\m$ are much smaller than the energy $E$. Propagating through the matter, jets get modified interacting with the medium-induced color field, and the momentum distribution is also affected. In the perturbative regime, the corresponding change in the momentum distribution can be studied order by order in the coupling $g$. One should also distinguish the coupling entering the in-medium potential and the emission vertex involving the energetic parton, see e.g. \cite{Sievert:2018imd}. 

After the amplitude is obtained, it should be squared and averaged over quantum numbers before one can construct the final momentum distribution. This procedure requires one to further specify the medium model by defining the multipoint correlations of external fields. Following the prescription commonly used in the pQCD considerations of the jet-medium interactions, we treat the medium color field to be classical and stochastic. We also assume a color neutrality condition requiring that 
$$
\left\langle t^a_i t^b_j \right\rangle = 
\frac{1}{d_\mathrm{tgt}} \tr\left( t^a_i t^b_j \right) 
=\frac{1}{2  C_{\bar{R}}} \delta_{i j} \delta^{a b}\,,
$$
and take into account only pairwise averages, see e.g. \cite{Sadofyev:2021ohn}. This approximation is motivated by the fact that interference terms are suppressed in a random classical system. We leave the color representation of the sources free, but assume that all of them are in the same representation. Here $d_{tgt}$ is the dimension of the color representation of the sources ("target"), and $C_{\bar R}$ is the quadratic Casimir of the opposite representation. The amplitude squared reads 
\begin{align}
\label{Msquared}
    \left\langle|M|^2\right\rangle &= \underbrace{\left\langle|M_0|^2\right\rangle}_{N=0}+\underbrace{\left\langle|M_1|^2\right\rangle+\left\langle M_2M_0^*\right\rangle+\left\langle M_0M_2^*\right\rangle}_{N=1}\notag\\
    &+\underbrace{\left\langle|M_2|^2\right\rangle+\left\langle M_3M_1^*\right\rangle+\left\langle M_1M_3^*\right\rangle+\left\langle M_4M_0^*\right\rangle+\left\langle M_0M_4^*\right\rangle}_{N=2}+...\:,
\end{align}
where we have identified the first several orders in the opacity expansion numerated with $N$, expressing them through the terms in the perturbation series $M (p)=\sum\limits_r M_r$ with $r$ counting the number of the in-medium field insertions. One should notice that the contributions to $\left\langle|M|^2\right\rangle$ involving an odd number of external fields average to zero under our assumptions.
\begin{figure}
    \vspace{-1cm}
    \centerline{
    \includegraphics[scale=0.23]{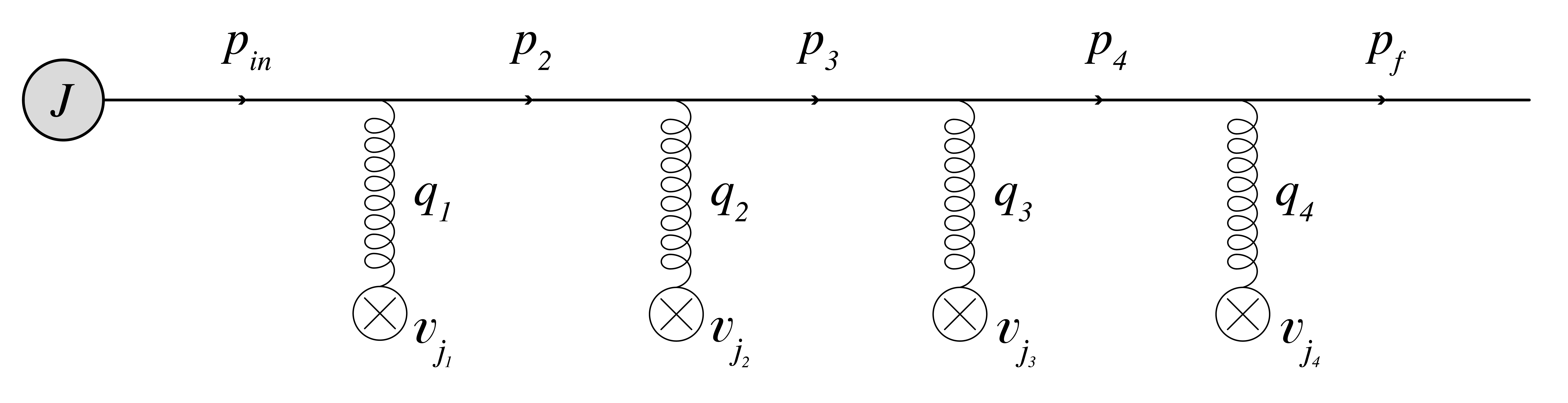}}
    \vspace{-0.5cm}
    \caption{The contribution $M_{r}$ with four field insertions ($r=4$) to the full amplitude.}
    \label{fig:Mr}
\end{figure}
The contribution $M_r$ to the full amplitude with $r$ external field entries, see Fig.~\ref{fig:Mr}, can be written as
\begin{align}
\label{Mrd3q}
    iM_r (p_f) &= \prod_{n=1}^{r}\Bigg[(-1)\sum_{j_n}\int \frac{d^2 \tvec{q}_n\, dq_{n,z}}{(2\pi)^3}\, t^a_\mathrm{proj} t_{j_n}^a\notag\\
    &\hspace{3cm}\times e^{-i \left(\tvec{q}_n \cdot \tvec{x}_{j_n}+q_{n,z} z_{j_n}\right)}\,\frac{2 E}{ p_n^2 + i \epsilon}\,
    v_{j_n} (q_n)\Bigg]J\left(p_{in}\right) \, ,
\end{align}
where $t^a_{proj}$ is the parton ("projectile") color generator, $p_n=p_f-\sum\limits_{m=n}^{N} q_m$, and $p_{in}=p_1$ under this numeration. The sums and integrals should be understood as acting on the whole expression, including $J\left(p_{in}\right)$ which depends on the momentum transfers $q_n$.

Assuming for a moment that all the source positions are different, we can perform the $q_z$ integrals by residues, noticing that only the small poles of the propagators contribute. Indeed, the large $q_z-$poles scale as $E$, and when one of such residues is substituted into the integrand that leads to a sub-eikonal result going beyond our accuracy. Similarly to the original GLV formalism, we also assume that the potentials $v_j(q)$ are screened at some scales $\m_j$ (varying from point to point), resulting in exponentially suppressed contributions for a sufficiently large and dilute medium $\m_{j_n} (z_{j_n}-z_{j_{n-1}})\gg1$. Then,
\begin{align}
\label{MN}
    iM_r &= \prod_{n=1}^{r}\Bigg[\sum_{j_n}\int \frac{d^2 \tvec{q}_{n}}{(2\pi)^2}\, it^a_\mathrm{proj} t_{j_n}^a\,\theta_{j_n,j_{n-1}}\notag\\
    &\hspace{3cm}\times e^{-i \tvec{q}_{n} \cdot \tvec{x}_{j_n}}e^{-iQ_{n}\left(z_{j_n}-z_{j_{n-1}}\right)}\,
    v_{j_n} (q_n)\Bigg]J\left(E, \tvec{p}_{in}\right) \, ,
\end{align}
where $J\left(E, \tvec{p}_{in}\right)$ is the eikonal limit of the source function with $p_{in}$ substituted from the corresponding pole, $\theta_{j_n,j_{n-1}}\equiv\theta\left(z_{j_n}-z_{j_{n-1}}\right)$, the Landau-Pomeranchuk-Migdal (LPM) phase $Q_n\equiv\frac{p_{n\perp}^2-p_{f\perp}^2}{2E}$ comes from the small pole of the propagator $p_n^2=0$, and without loss of generality we set $z_{i_0}=0$, which could be thought of as the center of the source function. Notice that while the LPM phases are sub-eikonal, they are enhanced by the large medium length, and, thus, should be kept explicitly, {\it c.f.} \cite{Sievert:2018imd, Sadofyev:2021ohn}.

Now one has to square the full amplitude and average over quantum numbers to obtain the momentum distribution modified by the medium. However, looking at (\ref{Msquared}) one should notice that two fields in the pair can come from the same side of the cut. Then, due to the color neutrality, the general expression (\ref{MN}) involves one or more $\theta(0)$ which should be defined. Thus, we have to re-consider the derivation in the case when the amplitude involves two consequent interactions on the same source, or, in other words, we have to separately study the so-called double-Born (DB) diagrams in addition to the ones involving only direct single-Born (SB) interactions. The former are needed to ensure unitarity. If such a contact interaction takes place, involving two consequent insertions $n$ and $n+1$ in $M_r$, then the color averaging results in $j_n=j_{n+1}$, and the corresponding $q_z$ integrals read
\begin{align}
\label{DBgeneral}
&(4E^2)\int \frac{dq_{n,z}}{2\pi}\frac{dq_{n+1,z}}{2\pi}\frac{v_{j_n}(q_n)v_{j_n}(q_{n+1})}{\left(p_n^2+i\e\right)\left(p_{n+1}^2+i\e\right)}e^{-iq_{n,z}\left(z_{j_n}-z_{j_{n-1}}\right)}e^{-iq_{n+1,z}\left(z_{j_n}-z_{j_{n-1}}\right)}\simeq
\notag\\
&\hspace{1cm}-2i E\,\theta_{j_n,j_{n-1}}\,e^{-i\left(q^{(p)}_{n,z}+q_{n+1,z}\right)\left(z_{j_n}-z_{j_{n-1}}\right)}\,\int\frac{dq_{n+1,z}}{2\pi}\frac{v_{j_n}\left(\tvec{q}_{n},q^{(p)}_{n,z}\right)v_{j_n}(\tvec{q}_{n+1},q_{n+1,z})}{p_{n+1}^2+i\e}
\notag\\
&\hspace{2cm}\simeq-\frac{1}{2}\,\theta_{j_n,j_{n-1}}\,v_{j_{n}}(q^2_{n\perp})v_{j_{n}}(q^2_{n+1\perp})\,e^{-i\left(q^{(p)}_{n,z}+q_{n+1,z}\right)\left(z_{j_n}-z_{j_{n-1}}\right)}
\end{align}
where $q^{(p)}_{n,z}=Q_n-\sum\limits_{m=n+1}^N q_{z,m}$ is the small pole of $p_n^2=0$ solved for $q_{n,z}$, the combination $q^{(p)}_{n,z}+q_{n+1,z}$ is $q_{n+1,z}$-independent, and the final expression is written under an assumption that $\sum\limits_{m=n+2}^N q_{z,m}$ is sub-eikonal, as is the case after the $q_{n+2,z}$ integration, for additional details see Appendix~\ref{App1}. It should be also mentioned that the full $q_{n+2,z}$-dependence of the integrand results only in additional screened poles, and the corresponding integration is unaffected by the presence of the contact interaction. Finally, one should notice that we have omitted the $q_{n,z}$-dependent entry of the form $v\left(\tvec{q}_{n-1},q^{(p)}_{n-1,z}\right)$ in the integrand since it cannot modify the $q_{n,z}$ integration, and after the first integration over $q_{n,z}$ its $q_{n+1,z}$-dependence disappears. Each SB contribution to the amplitude squared should be supplemented with all the corresponding contact terms.

When the amplitude is squared, each contribution to the $N$th order in opacity involves $2N$ sums over the in-medium sources. The color averaging reduces the number of the sums to $N$: only two gluon exchanges are allowed, happening either on different sides of the cut (SB interactions) or on the same source on one of the sides (DB interactions). Commonly, at this step, the discrete sums are replaced by continuous averages with a source number density, i.e.
\begin{equation}\label{eq:averaging}
\sum_i f_i = \int d^2 \tvec{x}\,dz \: \rho(\tvec{x},z) \: f(\tvec{x},z)\, , 
\end{equation}
where the spatial integration goes over the medium volume, and we assume that the medium is large but keep its finite longitudinal length $L$ explicitly. 

If the system is uniform in the transverse directions, each $\tvec{x}_{n}$ integral acts only on the corresponding Fourier factor, resulting in
$$
    \int d^2 \tvec{x}_{n} \, e^{- i (\tvec{q}_{n}\pm\overline{\tvec{q}}_{n}) \cdot \tvec{x}_{n}} =(2\pi)^2\,\d^{(2)}(\tvec{q}_{n}\pm\overline{\tvec{q}}_{n})
$$
where $\tvec{q}_n$ and $\overline{\tvec{q}}_n$ are the two momentum exchanges in the averaged potential pair and the sign is different for SB and DB interactions. Thus, the number of the transverse momentum integrals is halved, while all the LPM phases cancel out in the jet momentum distribution. 

Collecting the SB and DB terms, one finds the well-known result \cite{Gyulassy:2002yv} for the amplitude squared at $N$th order in opacity:
\begin{align}
\label{MR0}
    \left\langle \left| M \right|^2 \right\rangle^{(N)} &=
    \prod_{n=1}^N\left[(-1)\int\limits_{0}^{z_{n+1}} dz_n \int \frac{d^2\tvec{q}_{n}}{(2\pi)^2} \,\mathcal V(\tvec{q}_n,z_n)
    \right] \: 
    \left|J\left(E,\tvec{p}_{in}\right)\right|^2\,
\end{align}
with 
$$
\mathcal V(\tvec{q},z)\equiv -\mathcal{C}\,\r(z)\left(\left|v(q_\perp^2)\right|^2-\delta^{(2)}(\tvec{q})\int d^2\tvec{l}\,\left|v(l_\perp^2)\right|^2\right)\,,
$$
where $\mathcal V(\tvec q,\,z)$ is a specific combination of the in-medium color potentials, which enters the distribution at all orders in opacity, and $z_{N+1}=L$. We will refer to it as the dipole potential, since it can be related to the forward scattering amplitude for a color dipole. Here, $\mathcal{C}=\frac{\mathcal{C}_{\text{proj}}}{2C_{\bar{R}}}$ is the full color factor with $\mathcal{C}_{\text{proj}}\tvec{1}=t^a_{\text{proj}}t^a_{\text{proj}}$, and the superscript $(N)$ indicates the order in the opacity expansion. Notably, the opacity series can be now resummed, since the convolution (\ref{MR0}) reduces to a local product in the $\tvec x$-space. Indeed, introducing the jet distribution in the $\tvec x$-space
\bea
E\frac{d\N}{d^2\tvec{x}dE}=\frac{1}{2(2\pi)^3}\int\,\frac{d^2\tvec{p}}{(2\pi)^2}\,\left\langle \left| M(\tvec{p}) \right|^2 \right\rangle e^{i\tvec{p}\cdot\tvec{x}}
\eea
we readily write
\bea
\label{sumGLV0}
\frac{d\N}{d^2\tvec{x}dE}=\sum_{N=0}^{\infty}\int\,\frac{d^2\tvec{p}\,d^2\tvec{r}}{(2\pi)^2}e^{i\tvec{p}\cdot\left(\tvec{x}-\tvec{r}\right)}\,\frac{(-1)^N\left[\mathcal{V}(\tvec{r})L\right]^N}{N!}\frac{d\N^{(0)}}{d^2\tvec{r}dE}=e^{-\mathcal{V}(\tvec{x})L}\frac{d\N^{(0)}}{d^2\tvec{x}dE}
\eea
where for simplicity we set $\r(z)=const$, and, consequently, $\mathcal{V}(\tvec q, z)=\mathcal{V}(\tvec q)$. 

On the other hand, if the medium is inhomogeneous in the transverse directions, then the $\tvec{x}_{n}$ integrals cannot be simplified without further assumptions. As in \cite{Sadofyev:2021ohn} we focus on the leading corrections in the case of a slow $\tvec{x}-$dependence, when the thermodynamic parameters can be expanded in their transverse gradients. Then, the leading gradient corrections to the general transverse integral appear due to linear terms, such as
$$
    \int d^2 \tvec{x}_{n} \,x^\a_{n}\, e^{- i (\tvec{q}_{n}\pm\overline{\tvec{q}}_{n}) \cdot \tvec{x}_{n}} =i\,(2\pi)^2\,\frac{\pa}{\pa(q_{n}\pm\overline{q}_{n})_\a}\,\d^{(2)}(\tvec{q}_{n}\pm\overline{\tvec{q}}_{n})
$$
where $\a$ is a vector index in the transverse 2D space.

In the absence of medium flow, the two medium parameters of interest are $\r$ and $\m$, and to the leading order in gradients they can be written as
\begin{align}\label{eq:grad_exp}
    &\rho(\x, z) \approx \rho(z) + \tvec\na\rho(z) \cdot \x \, , \nn
    &\mu^2 (\x, z) \approx \mu^2(z) + \tvec\na\mu^2(z) \cdot \x  \, , 
\end{align}
where for compactness we use $\r(z)\equiv\r(0,z)$. Integrating the $\d$-function derivatives by parts, one may find that they act only on the LPM phases, while all other contributions are either suppressed within the eikonal expansion or cancel between complex conjugated contributions. This generalizes the observation in \cite{Sadofyev:2021ohn} for the broadening at the first order in opacity. We will discuss the details of this derivation in a separate Appendix~\ref{App1} using the $N=2$ case as an example. Here we only present the squared amplitude at $N$th order in opacity:
\begin{align}
\label{MR1}
\left\langle \left| M\right|^2 \right\rangle^{(N)} & = \prod\displaylimits_{n=1}^N\left[\int\displaylimits_0^{z_{n+1}}\, dz_n\int \frac{d^2\tvec{q}_n}{(2\pi)^2}\right] 
 \left(1+ \frac{1}{E}\sum_{m=1}^N (z_m-z_{m-1})\tvec{p}_m \cdot \sum_{k=m}^N
\hat{\tvec{g}}_k\right)\nn &\hspace{1cm}\times(-1)^N\mathcal{V}_1(\tvec{q}_{1})\,...\, \mathcal{V}_N(\tvec{q}_{N})|J(E, \tvec{p}_{in})|^2 \, , 
\end{align}
where we have introduced notations $\r_k$, $\m_k$, and $\mathcal{V}_k$ to distinguish different sources of gradients,  $\hat{\tvec{g}}_k\equiv\left(\tvec\na\r\frac{\d}{\d\r_k}+\tvec\na\m^2\frac{\d}{\d\m^2_k}\right)$ is an operator generating gradient contributions, and the ordering of the $z$-integrals is enforced by the $\theta_{n,n-1}$ in (\ref{MN}) and (\ref{DBgeneral}). After the variations with respect to $\r_k$ and $\m_k$ are performed, we again set the thermodynamic parameters of the same type to be equal and constant in $z$. 

Further simplifying (\ref{MR1}) and transforming to the $\tvec{x}$-space, we find
\begin{align}
\frac{d\N^{(N)}}{d^2\tvec{x}dE} & = \int\, \frac{d^2\tvec{p}\,d^2\tvec{r}}{(2\pi)^2}\,
 e^{i\tvec{p}\cdot\left(\tvec{x}-\tvec{r}\right)}(-1)^N\big[\mathcal{V}(\tvec{r})L\big]^N\Bigg\{\frac{1}{N!} + \frac{L}{E(N+1)!}\notag\\
 &\times\sum_{m=1}^N\Bigg[(N+1-m)\p\cdot\left(\frac{\mathcal{V}'(\tvec{r})}{\mathcal{V}(\tvec{r})}\tvec\na\mu^2+\frac{1}{\r}\tvec\na\r\right)+i(N+1-m)^2\frac{\tvec\na\mathcal{V}(\tvec{r})}{\r\,\mathcal{V}(\tvec{r})}\cdot\tvec\na\r\notag\\
 &+i(N+1-m)\left(\frac{\tvec\na\mathcal{V}'(\tvec{r})}{\mathcal{V}(\tvec{r})}+(N-m)\frac{\mathcal{V}'(\tvec{r})}{\mathcal{V}(\tvec{r})}\frac{\tvec\na\mathcal{V}(\tvec{r})}{\mathcal{V}(\tvec{r})}\right)\cdot\tvec\na\mu^2\Bigg]\Bigg\}\frac{d\N^{(0)}}{d^2\tvec{r}dE}\,,
\end{align}
and the opacity series can be again resummed, resulting in
\begin{align}\label{eq:compare_to_BDMPS}
\frac{d\N}{d^2\tvec{x}dE} & = e^{-\mathcal{V}(\tvec{x}) L}\Bigg\{\left[1-i\frac{\mathcal{V}(\tvec{x})L^3}{6E}\Bigg(\frac{\mathcal{V}'(\tvec{x})}{\mathcal{V}(\tvec{x})}\tvec\na\mu^2+\frac{1}{\r}\tvec\na\r\Bigg)\cdot\tvec\na\mathcal{V}(\tvec{x})\right]\frac{d\N^{(0)}}{d^2\tvec{x}dE}\notag\\
 &+i\frac{\mathcal{V}(\tvec{x})L^2}{2E}\left(\frac{\mathcal{V}'(\tvec{x})}{\mathcal{V}(\tvec{x})}\tvec\na\mu^2+\frac{1}{\r}\tvec\na\r\right)\cdot\tvec\na\frac{d\N^{(0)}}{d^2\tvec{x}dE}\Bigg\}\,,
\end{align}
where $\mathcal{V}'(\tvec{x})=\frac{\pa}{\pa\mu^2}\mathcal{V}(\tvec{x})$. Thus, we have derived \eqref{eq:global}, which is one of the main results of this work. It gives the Fourier transform of the momentum broadening distribution up to the first order in gradients and to all orders in opacity. One can further use it to study the jet momentum broadening, which we proceed to do in Section~\ref{ref:pheno}.

\section{The BDMPS-Z Formalism}\label{ref:BDMPS}

In this section, we re-derive the leading gradient effects on the jet momentum broadening within the BDMPS-Z approach. From a practical point of view, in this formalism the resummation of multiple field insertions is first performed at the amplitude level by constructing the dressed in-medium propagator. We obtain the in-medium propagator for an inhomogeneous medium, including the leading gradient contributions.

Since the interactions with the medium are dominated by tree-level gluon exchanges, the dynamics of the background field is dominated by the classical Yang-Mills equations, and it can be treated as a stochastic variable. In the BDMPS-Z approach, it is typically assumed that the statistics of the field take a white-noise form due to the large number of uncorrelated degrees of freedom in the medium, analogous to the McLerran-Venugopalan model~\cite{Mehtar-Tani:2006vpj,McLerran:1993ka,McLerran:1993ni}. This is equivalent to the assumption used in the previous section that only pairwise averages are non-negligible. 

Thus, we again start with a model for the in-medium color field. The model commonly used in the BDMPS-Z approach can be conveniently summarized with
\begin{align}
\label{e:potl2}
    g A_\mathrm{ext}^{\mu a} (q) = (2\pi)\,g^{\m 0}\,v^a(q)
    \: \delta\left(q^0\right) \,,
\end{align}
where the scattering potential should be set to $v^a(q)=\sum_j e^{-i( \tvec{q} \cdot \tvec{x}_j+ q_z z_j)}\, t_j^a\,v_j(q)$ to coincide with the GW model used in the previous section. It should be also noticed that in general the individual potentials of the scattering centers can be left unspecified in both approaches, although one should be careful treating the DB interactions. 

Having the form of the scattering potential, we proceed to rewrite the amplitude given in \eqref{MN} in terms of an effective dressed propagator. For that, we first write \eqref{Mrd3q} using the in-medium color field \eqref{e:potl2}, then
\begin{align}
\label{Mp3D}
    iM_r (p_f) = \prod_{n=1}^{r}\left[(-1)\int \frac{d^2 \tvec{p}_n\,dp_{n,z}}{(2\pi)^3}\,t^a_\mathrm{proj} v^a(p_{n+1}-p_{n})\,\frac{2E}{p_n^2+i\e}\right]\,J\left(p_{in}\right) \, ,
\end{align}
where the temporal components satisfy the constraint $p^0_n=E$, indicating that there is no energy transfer between the medium and the probe via soft gluon exchanges. It is convenient to Fourier transform the potentials, simplifying the $p_z$-integration. This is equivalent to working in a mixed representation, where "time" dependence is made explicit, commonly employed in the BDMPS-Z related literature and analogous to old-fashioned perturbation theory. Then, \eqref{Mp3D} can be rewritten as
\begin{align}
\label{Mp2D}
    iM_r (p_f) &= \prod_{n=1}^{r}\Bigg[\int \frac{d^2 \tvec{p}_n\, d^2\tvec{x}_n\,dz_n}{(2\pi)^2}
    \,\theta_{n,n-1}\,i t^a_\mathrm{proj} v^a(\tvec{x}_n,z_n)\notag\\
    &\hspace{2cm}\times e^{-i(\tvec{p}_{n+1}-\tvec{p}_{n})\cdot \tvec{x}_n}e^{-i\frac{p_{n\perp}^2}{2E}(z_n-z_{n-1})}\Bigg]\,e^{i\frac{p_{f\perp}^2}{2E}z_{r}}\,J\left(E,\tvec{p}_{in}\right) \, ,
\end{align}
where $(\tvec{x}_n, z_n)$ denote the interaction points, and should not be mixed with the source coordinates in the GW model, now hidden in $v^a$.

Finally, noticing that $\p_1=\p_{in}$ and $\p_{r+1}=\p_f$, we can write the perturbative amplitude as a 
convolution between the source and a contribution to an effective 
single particle propagator $G_r(\tvec{p}_{f},L;\tvec{p}_{in},0)$:
\begin{align}
\label{e:BroadBorn2.1}
    iM_r (p) = \int \frac{d^2 \p_{in}}{(2\pi)^2}e^{i\frac{\tvec{p}_{f}^2}{2E}L}\,G_r(\tvec{p}_{f},L;\tvec{p}_{in},0)\,J\left(E,\tvec{p}_{in}\right) \, .
\end{align}
Notice that the in-medium potential $v^a$ is screened by the Debye mass $\mu$, and thus has a finite 
spatial support of size $\sim\frac{1}{\m}$. As a consequence, the $z$-integrals can be safely taken to run from the production point $z=0$ to the end of the medium at $z=L$, which enters the single particle propagator.

The full effective propagator can be obtained by summing over the number of interactions
\begin{align}
\label{e:BroadBorn2.2}
    G(\tvec{p}_{f},L;\tvec{p}_{in},0)=\sum_{r=0}^\infty G_r(\tvec{p}_{f},L;\tvec{p}_{in},0)\, ,
\end{align}
where it is simple to check that in the case of vacuum propagation it reduces to the usual Feynman result
\begin{align}
    G_0(\tvec{p}_{f},L;\tvec{p}_{in},0)=(2\pi)^2\delta^{(2)}(\tvec{p}_{f}-\tvec{p}_{in})e^{-i\frac{p_{f\perp}^2}{2E}L} \, .
\end{align}
Inserting this back in \eqref{e:BroadBorn2.1}, we find that the amplitude reduces to the initial source function, as expected. 

Instead of dealing with the series in \eqref{e:BroadBorn2.2}, one can construct an evolution equation for $G$ from \eqref{Mp2D} and \eqref{e:BroadBorn2.1}, which takes the usual Schr\"{o}dinger-like form
\begin{align}
\label{ShEq}
    \frac{\pa}{\pa L}G(\tvec{p}_{f},L;\tvec{p}_{in},0)&=-i\frac{p_{f\perp}^2}{2E}G(\tvec{p}_{f},L;\tvec{p}_{in},0)\notag\\
    &\hspace{-1cm}+\int\frac{d^2\tvec{l} d^2\tvec{x}}{(2\pi)^2}\,it_{\text{proj}}^a\,v^a(\x,L)\,e^{-i(\tvec{p}_f-\tvec{l})\cdot\x}\,G(\tvec{l},L;\tvec{p}_{in},0)\,.
\end{align}
This equation should be supplemented with an "initial condition", which can be obtained from the fact that at $L=0$ there is no modification to the amplitude sourced by $J(p_{in})$. The solution to (\ref{ShEq}) with the corresponding initial condition is well known~\cite{book:Kleinert_path_integrals}, and its $\tvec x$-space form can be written as a path integral
\begin{align}\label{eq:G_prop}
 G(\x_L,L;\x_0,0)&= \int\limits_{\x_0}^{\x_L} \mathcal D\rn  \exp\left(\frac{iE}{2}\int\limits_{0}^{L}  d\t \, \dot{\rn}^2\right)\notag\\
 &\hspace{1cm}\times\cP\exp\left(i\int\limits_{0}^{L}  d\t \, t^a_{\text{proj}}v^a(\rn(\t),\t)\right) \,,  
\end{align}
where $\cP$ indicates path ordering, and $\tvec{x}_L\equiv\tvec{x}(L)$ and $\tvec{x}_0\equiv\tvec{x}(0)$ are the boundary conditions for the trajectory. This effective propagator can be thought of as describing a massive non-relativistic particle, moving from the initial position $\x_0$ at "time" $\t=0$ to the final position $\x_L$ at "time" $\t=L$ in a (random) potential $v^a(\tvec{r},\t)$.

With the further assumption that the QCD emission vertices are unaltered in the medium, one can derive a set of effective Feynman rules using the propagator above and compute any quantum amplitude. As a consequence, in such a path integral formulation of the BDMPS-Z formalism, in practice one can just draw all the relevant time ordered Feynman diagrams including the medium and directly obtain the amplitudes, similar to more standard vacuum pQCD calculations. The squared amplitude, already averaged over the quantum numbers and medium configuration, can be easily expressed through an in-medium correlation function of two propagators, {\it c.f.} \eqref{e:BroadBorn2.1},
\begin{align}\label{eq:M2_GG}
 \langle|M|^2\rangle=\int \frac{d^2\p_{in}d^2\overline{\p}_{in}}{(2\pi)^4}\langle G(\p_{f},L;\p_{in},0)G^\dagger(\p_{f},L;\overline{\p}_{in},0)\rangle J(E,\p_{in})J^*(E,\overline{\p}_{in})\,,
\end{align}
as well as the distribution corresponding to the jet momentum broadening itself.

In order to compute \eqref{eq:M2_GG}, we first have to revisit how the averaging procedure is performed in the BDMPS-Z approach. 
Since $G(\x_L,L;\x_0,0)$ is 
a functional 
of $v^a(\tvec{r}(\t),\t)$
, one first needs to consider the average of the in-medium color fields. Since these are assumed to have Gaussian statistics, only two-point functions of the potentials are non-trivial. For instance, in the GW model, the corresponding average reads 
\begin{align}\label{vvint}
&\langle t^a_{\text{proj}}v^a(\rn,\t)t^b_{\text{proj}}v^{\dagger b}(\overline\rn,\overline\t)\rangle=\mathcal{C}\,g^4\, \int dz\,d^2\tvec x \, \r(\tvec x,\,z)\notag\\
&\hspace{1cm} \times \int \frac{d^2\q\,dq_z\: d^2\overline\q\,d\overline{q}_z}{(2\pi)^6} \frac{e^{i\q\cdot\left(\tvec{r}-\tvec{x}\right)}e^{-i\overline\q\cdot \left(\overline{\tvec{r}}-\tvec{x}\right)}e^{iq_z\left(\t-z\right)}e^{-i\overline{q}_z\left(\overline\t-z\right)}}{(q_\perp^2+q_z^2+\m^2(\tvec{x},z))(\overline{q}_\perp^2+\overline{q}^2_z+\m^2(\tvec{x},z))}\, ,
\end{align}
where we have used the color neutrality condition and taken the continuous limit of the distribution of scattering centers in the medium (as in the previous section) in order to make the dependence on the source density explicit. Before we proceed, one should notice that only a particular limit of this correlation function enters the amplitude (\ref{Mp2D}) when it is squared and averaged, see e.g. \cite{Blaizot:2012fh}. In fact, the average in (\ref{vvint}) is Fourier transformed to the momentum space in the amplitude squared, with the $z$-component of the momentum being sub-eikonal. Thus, the $q_z$ dependence in (\ref{vvint}) can be neglected without affecting the final result at the accuracy level being considered. Using this simplification, we find
\begin{align}\label{eq:AA_2}
\langle t^a_{\text{proj}}v^a(\rn,\t)t^b_{\text{proj}}v^{\dagger b}(\overline\rn,\overline\t)\rangle &\simeq \left(1+\frac{\rn(\t)+\overline\rn(\t)}{2}\cdot\hat{\tvec{g}}\right)\notag\\
&\hspace{1cm}\times\mathcal{C}\,\d(\t-\overline\t)\,\r\,g^4\int \frac{d^2\tvec q}{(2\pi)^2} \frac{e^{i\tvec q\cdot(\rn-\overline\rn)}}{(\tvec{q}^2+\m^2)^2}\,,
\end{align}
where $\tvec x$ in $\rho(\tvec x,\,z)$ and $\m^2(\tvec x,\,z)$ has been replaced with $-i\frac{\pa}{\pa (\tvec q-\overline{\tvec q})}$ acting on everything but the delta function $\d^{(2)}(\tvec q-\overline{\tvec q})$, and we assume that $\r$ and $\m^2$ are constant in the longitudinal direction.
One should notice that all the previous steps in this section are unaffected by the inhomogeneity of the medium, and the gradient effects enter solely through the potential averages.

\begin{figure}[h!]
    \vspace{-1cm}
    \centering
    \includegraphics[scale=.6]{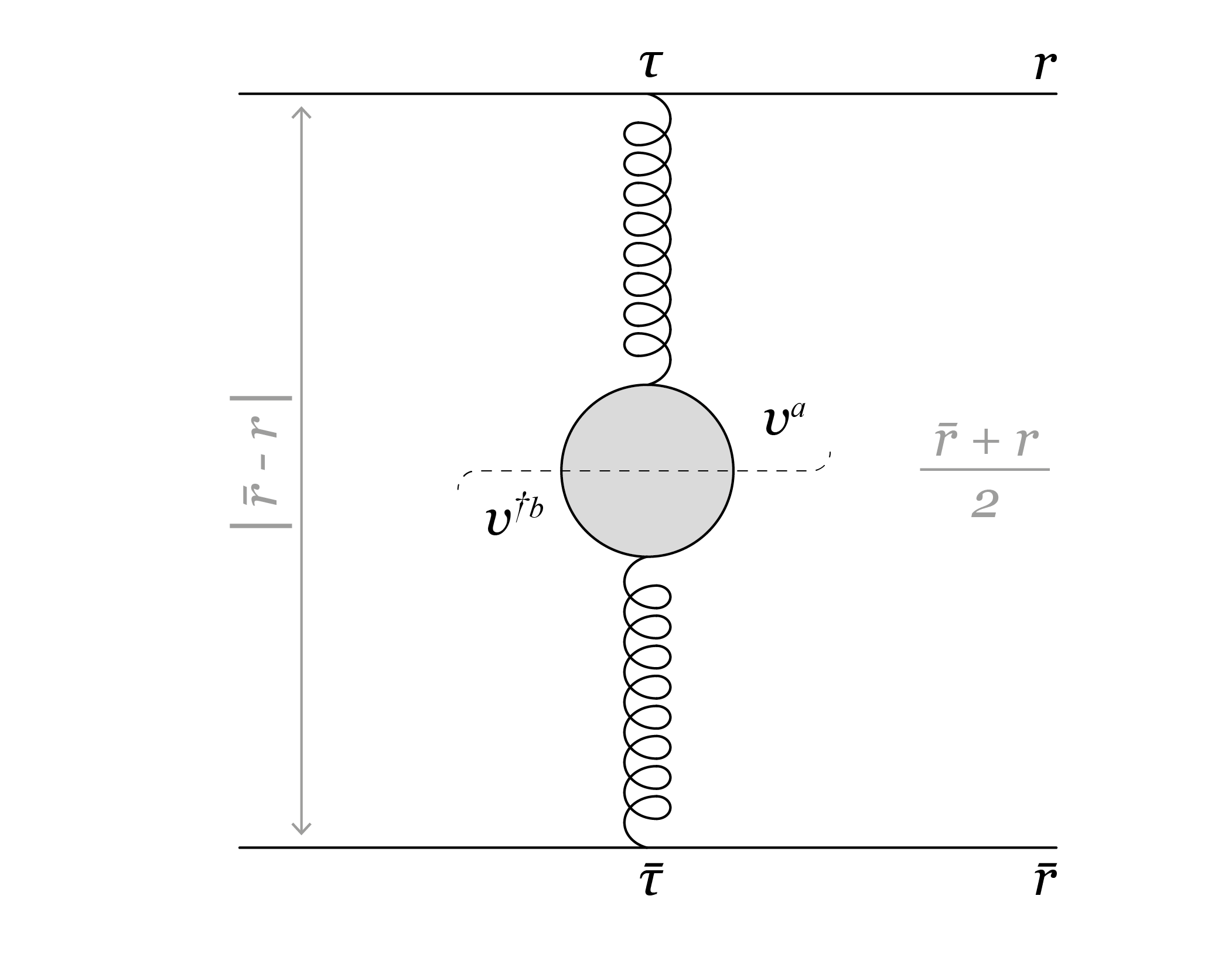}
    \vspace{-1cm}
    \caption{Diagrammatic illustration of \eqref{eq:AA_2}. Here, the amplitude level process is depicted in the upper half of the plot while the conjugate amplitude part is depicted in the bottom one, with the dashed cut in between.}
    \label{fig:cartoon_CM}
\end{figure}

The two-point correlator of the in-medium color potentials now depends not only on the transverse  
size of the effective color dipole formed by the propagating parton in amplitude and conjugate amplitude (see Fig.~\ref{fig:cartoon_CM}), which is proportional to the difference $|\rn-\overline\rn|$, but also on its center of mass transverse position $\frac{\rn+\overline\rn}{2}$. This is well expected, since translation invariance is now violated by the transverse gradients. The higher-order gradients will enter the BDMPS-Z construction similarly, with higher powers of the transverse position of the center of mass.

Given the leading gradient form for the pairwise average of two in-medium potentials, we now turn to the average entering \eqref{eq:M2_GG}. Using the two-point correlator \eqref{eq:AA_2}, we can write the average of two Wilson lines as\footnote{The exponentiation follows from the fact that the averaging statistics is still Gaussian.}
\begin{align}
&\left\langle \cP\exp\left(i\int_{0}^{L}  d\t \, t_{\text{proj}}^av^a(\rn(\t),\t)\right)\cP\exp\left(-i\int_{0}^{L}  d\overline\t \, t_{\text{proj}}^bv^b(\overline\rn(\overline\t),\overline\t)\right)\right\rangle \notag\\
&\hspace{1cm}=\exp\left\{-\int\limits_0^L\,d\t\,\left[1+\frac{\tvec{r}(\t)+\overline\rn(\t)}{2}\cdot\hat{\tvec{g}}\right]\mathcal{V}\left(\tvec r(\t)-\overline\rn(\t)\right)\right\}\,,
\end{align}
where the contact terms arise from the pairwise averages of potentials coming from the same exponential, and we use the fact that the potential is Hermitian in the GW model. Notice that the exponential with the gradients in the argument should be treated as a series valid up to the first order at the accuracy of our consideration. 

The position space form of the relevant correlator of two propagators reads
\begin{align}\label{eq:GG_p_space_redux}
 &\left\langle G(\x_L,L;\x_0,0)G^\dagger(\overline{\x}_L,L;\overline{\x}_0,0)\right\rangle\equiv\left\langle G(\x_L;\x_0)G^\dagger(\overline{\x}_L;\overline{\x}_0)\right\rangle\notag\\
 &\hspace{1cm}=\int\limits_{\x_0}^{\x_L} \mathcal D\rn \int\limits_{\overline{\x}_0}^{\overline{\x}_L} \mathcal D\overline\rn \, \exp\left\{\frac{iE}{2}\int_{0}^{L}  d\t \, \left[\dot{\rn}^2-\dot{\overline\rn}^2\right]\right\}\notag\\
 &\hspace{2cm}\times\exp\left\{-\int\limits_0^L\,d\t\,\left[1+\frac{\rn(\t)+\overline\rn(\t)}{2}\cdot\hat{\tvec{g}}\right]\mathcal{V}\left(\rn(\t)-\overline\rn(\t)\right)\right\}\notag\\
 &\hspace{1cm}=\int\limits_{\tvec{u}_0}^{\tvec{u}_L} \mathcal D \un \int\limits_{\tvec{w}_0}^{\tvec{w}_L} \mathcal D \tvec w \, \exp\left\{\int_{0}^{L}  d\t \,\left[ iE\,\dot\un\cdot\dot{\tvec{w}}-\left(1+\tvec w\cdot\hat{\tvec{g}}\right)\mathcal{V}\left(\tvec u(\t)\right)\right]\right\}\,,
\end{align}
where $\tvec u\equiv\tvec r-\overline\rn$ and $\tvec w\equiv\frac{\tvec r+ \overline\rn}{2}$. The path integral is simple enough to be evaluated analytically, since the $\tvec w$-dependence enters only through the small gradient terms. At the first order in gradients the final answer is proportional to the value of the integrand on the solution of the effective classical equation of motion\footnote{One should notice that this EOM involves a complex potential term, which in fact should be understood as a perturbation around the trivial trajectory, see Appendix~\ref{App2}.} (EOM). 
Thus, using the standard methods~\cite{book:Kleinert_path_integrals,Blaizot:2012fh, Apolinario:2014csa}, we can reduce this expression to
\begin{align}
\label{GGfinal}
 &\left\langle G(\x_L;\x_0)G^\dagger(\overline{\x}_L;\overline{\x}_0)\right\rangle
 =\left(\frac{E}{2\pi L}\right)^2\,\frac{\exp\left\{iE\left( \tvec{w}\cdot\dot{\un}_c\right)\Big|_0^L-\int\limits_0^L\,d\t\,\mathcal{V}\left(\tvec{u}_c(\t)\right)\right\}}{1+\frac{i}{EL}\hat{\tvec{g}}\cdot\int\limits^L_0d\zeta\int\limits^\zeta_0d\xi\,\xi\,\tvec{\na}\mathcal{V}\left(\un_c(\xi)\right)} \,  ,
\end{align}
where $\tvec{u}_c(\t)$ is the classical solution satisfying 
\begin{align}\label{eq:ode}
&E\ddot\un=i\hat{\tvec{g}}\,\mathcal{V}(\un(\t))\, ,
\end{align}
with appropriate boundary conditions. Notice that the structure in the denominator of \eqref{GGfinal} is a part of a Jacobian appearing in the path integral, along with the overall multiple. In general, \eqref{eq:ode} does not admit an analytical solution.
However, since we are only interested in the leading gradient terms, it 
can be solved in an expansion $\un_c=\un_c^{(0)}+\un_c^{(1)}$, where $\un_c^{(0)}$ has no gradient dependence and $\un_c^{(1)}$ is linear in gradients\footnote{Notice that the imaginary part of the classical EOM is proportional to the gradient terms. Thus, substituting $\tvec{u}_c(\t)$ into the integrand is equivalent to acting on it with a shift operator changing the zero acceleration of the leading order trajectory with a complex function.}.

At the zeroth order, the right-hand side of the EOM is zero, and one readily finds that the separation vector can only change linearly with "time"
\begin{align}\label{eq:u0}
&\un^{(0)}_c(\t)=\frac{\un_L-\un_0}{L}\t+\un_0\,,
\end{align}
leading to the uniform broadening result in (\ref{sumGLV0}), see e.g. \cite{Blaizot:2012fh, Apolinario:2014csa}. In turn, the leading gradient correction to the trajectory is purely imaginary
\begin{align}
\un^{(1)}_c(\t)&=\frac{i}{E}\,\hat{\tvec{g}}\left\{\int\limits^\t_0\,d\z\,\int\limits_0^\z\,d\xi\, \mathcal{V}\left(\un^{(0)}_c(\xi)\right)-\frac{\t}{L}\int\limits^L_0\,d\z\,\int\limits_0^\z\,d\xi\, \mathcal{V}\left(\un^{(0)}_c(\xi)\right)\right\}\,,
\end{align}
and satisfies  
the trivial boundary conditions $\un^{(1)}_c(L)=\un^{(1)}_c(0)=\0$.

The momentum scale $\p_f$ in \eqref{eq:M2_GG} corresponds to a measured quantity and has to be matched in the two propagators. 
For this particular projection of the two-point correlator (\ref{GGfinal}) we find 
\begin{align}\label{eq:GG_some}
 &\langle G(\tvec{p}_f,L;\tvec{p}_{in},0)G^\dagger(\tvec{p}_f,L;\overline{\tvec{p}}_{in},0)\rangle\notag\\
 &\hspace{1cm}=\left(\frac{E}{2\pi L}\right)^2\int\,d^2\un_0 \,d^2\tvec{w}_0\,d^2\un_L \,d^2\tvec{w}_L\,e^{-i\tvec{p}_f\cdot \un_L}e^{i\un_0\cdot \frac{\tvec{p}_{in}+\overline{\tvec{p}}_{in}}{2}+i\tvec{w}_0\cdot\left(\tvec{p}_{in}-\overline{\tvec{p}}_{in}\right)}\notag\\
 &\hspace{2cm}\times\frac{\exp\left\{iE\left( \tvec{w}\cdot\dot{\un}_c\right)\Big|_0^L-\int\limits_0^L\,d\t\,\mathcal{V}\left(\tvec{u}_c(\t)\right)\right\}}{1+\frac{i}{EL}\hat{\tvec{g}}\cdot\int\limits^L_0d\zeta\int\limits^\zeta_0d\xi\,\xi\,\tvec{\na}\mathcal{V}\left(\un_c(\xi)\right)}\notag\\
 &\hspace{1cm}=\frac{(2\pi)^2}{L^2}\int\,d^2\un_0\,d^2\un_L\,e^{-i\tvec{p}_f\cdot \un_L}e^{i\un_0\cdot \frac{\tvec{p}_{in}+\overline{\tvec{p}}_{in}}{2}}\notag\\
 &\hspace{2cm}\times\frac{\d^{(2)}\left(\dot{\un}_c(L)\right)\d^{(2)}\left(\tvec{p}_{in}-\overline{\tvec{p}}_{in}-E\dot{\un}_c(0)\right)}{1+\frac{i}{EL}\hat{\tvec{g}}\cdot\int\limits^L_0d\zeta\int\limits^\zeta_0d\xi\,\xi\,\tvec{\na}\mathcal{V}\left(\un_c(\xi)\right)}\exp\left\{-\int\limits_0^L\,d\t\,\mathcal{V}\left(\tvec{u}_c(\t)\right)\right\}\,.
\end{align}
where again the delta functions of complex arguments should be understood as a shift operator (with small imaginary shift parameter) acting on the corresponding delta function of the (leading) real part of the argument. Contracting this with the initial source functions in \eqref{eq:M2_GG}, one can find the corresponding squared amplitude 
\begin{align}\label{eq:apphelp}
&\langle|M|^2\rangle\simeq\frac{1}{(2\pi L)^2}\int\,d^2\tvec{P}_{in}\,d^2\un_0 \,d^2\un_L\,e^{-i\tvec{p}_f\cdot \un_L}e^{i\tvec{P}_{in}\cdot\un_0}\notag\\
 &\hspace{4cm}\times\frac{\d^{(2)}\left(\dot{\un}_c(L)\right)\exp\left\{-\int\limits_0^L\,d\t\,\mathcal{V}\left(\tvec{u}_c(\t)\right)\right\}}{1+\frac{i}{EL}\hat{\tvec{g}}\cdot\int\limits^L_0d\zeta\int\limits^\zeta_0d\xi\,\xi\,\tvec{\na}\mathcal{V}\left(\un_c(\xi)\right)}\left|J\left(E,\tvec{P}_{in}\right)\right|^2\,,
\end{align}
where, as in the derivation of \eqref{MR0}, we have assumed that $J$ has at most a constant imaginary phase, as it is for a tree-level 2-to-2 process, and introduced the symmetric momentum $\tvec{P}_{in}=\frac{\tvec{p}_{in}+\overline{\tvec{p}}_{in}}{2}$. We have also used that the leading shift of $\tvec{p}_{in}-\overline{\tvec{p}}_{in}$ in $J\left(\tvec{p}_{in}\right)J\left(\overline{\tvec{p}}_{in}\right)$ is zero at the first order in gradients, since the first derivative of this symmetric function with respect to $\tvec{p}_{in}-\overline{\tvec{p}}_{in}$ is zero at $\tvec{p}_{in}=\overline{\tvec{p}}_{in}$. 

Thus, we can write the corresponding Fourier transformed distribution as
\begin{align}
&\frac{d\N}{d^2\tvec{x}dE}\simeq\frac{1}{L^2}\int\,d^2\un_0 \,d^2\un_L\,\d^{(2)}(\tvec x-\tvec{u}_L)\notag\\
 &\hspace{2cm}\times\frac{\d^{(2)}\left(\dot{\un}_c(L)\right)\exp\left\{-\int\limits_0^L\,d\t\,\mathcal{V}\left(\tvec{u}_c(\t)\right)\right\}}{1+\frac{i}{EL}\hat{\tvec{g}}\cdot\int\limits^L_0d\zeta\int\limits^\zeta_0d\xi\,\xi\,\tvec{\na}\mathcal{V}\left(\un_c(\xi)\right)}\frac{d\N^{(0)}}{d^2\tvec{u}_0dE}\,.
\end{align}
The velocity constraint at $\t=L$ allows to simplify the form of $\un_c$. In particular, in the case of vanishing matter gradients, it implies directly from \eqref{eq:u0} that the size of the effective dipole is preserved during its evolution~\cite{Blaizot:2015lma}. In the present case, 
this constraint implies
\begin{align}
&\frac{\un_L-\un_0}{L}+\frac{i}{E}\,\hat{\tvec{g}}\Bigg\{\int\limits_0^L\,d\xi\, \mathcal{V}\left(\un^{(0)}_c(\xi)\right)-\frac{1}{L}\int\limits^L_0\,d\z\,\int\limits_0^\z\,d\xi\, \mathcal{V}\left(\un^{(0)}_c(\xi)\right)\Bigg\}=0\,,
\end{align}
and one can use it to relate $\un_0$ and $\un_L$. 

At the leading order in gradients we find
\begin{align}
&\un_L-\un_0+\frac{iL^2}{2E}\,\hat{\tvec{g}}\,\mathcal{V}\left(\tvec{u}_L\right)\simeq 0\,.
\end{align}
and, consequently, $\un_c$ takes a remarkably simple form
\begin{align}
&\un_c(\t)=\un_L+\frac{i}{E}\,\hat{\tvec{g}}\,\mathcal{V}\left(\un_L\right)\left\{\frac{(\t-L)^2}{2}\right\}\,.
\end{align}
The leading order term is the well known constant solution. The net effect of including gradients is to change the effective trajectory of the dipole separation 
with a constant (imaginary) shift in the acceleration which is fixed by the form of $\mathcal{V}(\tvec{x})$. One should also notice that the delta function constraining the velocity at $\t=L$ has a non-trivial dependence on $\tvec{u}_0$ which enters the argument also through $\mathcal{V}\left(\un^{(0)}_c(\xi)\right)$, leading to an additional multiple:
\begin{align}
L^2\left[1-\frac{iL}{E}\hat{\tvec{g}}\cdot\Bigg\{\int\limits_0^L\,d\xi-\frac{1}{L}\int\limits^L_0\,d\z\,\int\limits_0^\z\,d\xi\Bigg\}\,\left(1-\frac{\xi}{L}\right)\,\tvec{\na}\mathcal{V}\left(\un^{(0)}_c(\xi)\right)\right]^{-1} \,.
\end{align}

Combining all the previous results and expanding in the smallness of gradients, we conclude that the $\tvec x$-space form of the jet distribution reads
\begin{align}
\label{dNBDMPSZ}
\frac{d\N}{d^2\tvec{x}dE}&\simeq\exp\left\{-\mathcal{V}\left(\tvec{x}\right)L\right\}\Bigg\{\left[1-\frac{iL^3}{6E}\tvec\na\mathcal{V}\left(\tvec{x}\right)\cdot\hat{\tvec{g}}\,\mathcal{V}\left(\tvec x\right)\right]\frac{d\N^{(0)}}{d^2\tvec{x}dE}\notag\\
&\hspace{2cm}+\frac{iL^2}{2E}\,\hat{\tvec{g}}\,\mathcal{V}\left(\tvec x\right)\cdot\tvec\na\frac{d\N^{(0)}}{d^2\tvec{x}dE}\Bigg\}\,,
\end{align}
and one can easily see that it exactly coincides with (\ref{eq:compare_to_BDMPS}), obtained from the opacity series resummation. It should be mentioned that the two measure factors coming from the path integral and velocity constraint cancel, resulting in the simpler expression above. Thus, we have shown that the usual BDMPS-Z formalism can be extended to inhomogeneous backgrounds to the leading order in
gradient expansion. Conceptually, the approach can be extended in a straightforward manner to include higher order gradient effects. However, we note that in such a situation the path integrals to be solved are technically more involved (e.g. the $\tvec w$ dependence in \eqref{eq:GG_p_space_redux} would no longer be linear).

Before we turn to the properties of the distribution (\ref{dNBDMPSZ}), it should be also mentioned that its extension to the case of $z$-dependent medium profile can be readily obtained. After a straightforward generalization of the derivation in the longitudinally homogeneous case, one finds
\begin{align}\label{eq:non-static-result}
&\frac{d\N}{d^2\tvec{x}dE}\simeq\exp\left\{-\int_0^L\,d\t\,\mathcal{V}\left(\tvec{x},\t\right)\right\}\notag\\
&\hspace{0.5cm}\times\left\{\rule{0cm}{1cm}\right.\left[1-\frac{i}{E}\int\limits_0^L\,d\t\,\tvec\na\mathcal{V}\left(\tvec{x},\t\right)\cdot\left(\int\limits_L^\t\,d\z\,\int\limits_0^\z\,d\xi+(L-\t)\int\limits^L_0\,d\xi\right)\hat{\tvec{g}}(\xi)\mathcal{V}\left(\tvec x,\xi\right)\right]\notag\\
&\hspace{1cm}+\frac{i}{E}\int\limits_{0}^{L} d\z \int\limits^{\z}_0 \, d\xi \,\hat{\tvec{g}}\left(\xi\right)\mathcal{V}\left(\tvec{x},\xi\right)\cdot\tvec\na\left.\rule{0cm}{1cm}\right\} \frac{d\N^{(0)}}{d^2\tvec x dE}\,,
\end{align}

where $\hat{\tvec{g}}\left(\t\right)=\left(\tvec\na\r(\t)\frac{\d}{\d\r}+\tvec\na\m^2(\t)\frac{\d}{\d\m^2}\right)$.

\section{The Final State Distribution and Its Properties}\label{ref:pheno}

\begin{figure}[t!]
    \vspace{-1cm}
    \centerline{
    \includegraphics[scale=0.42]{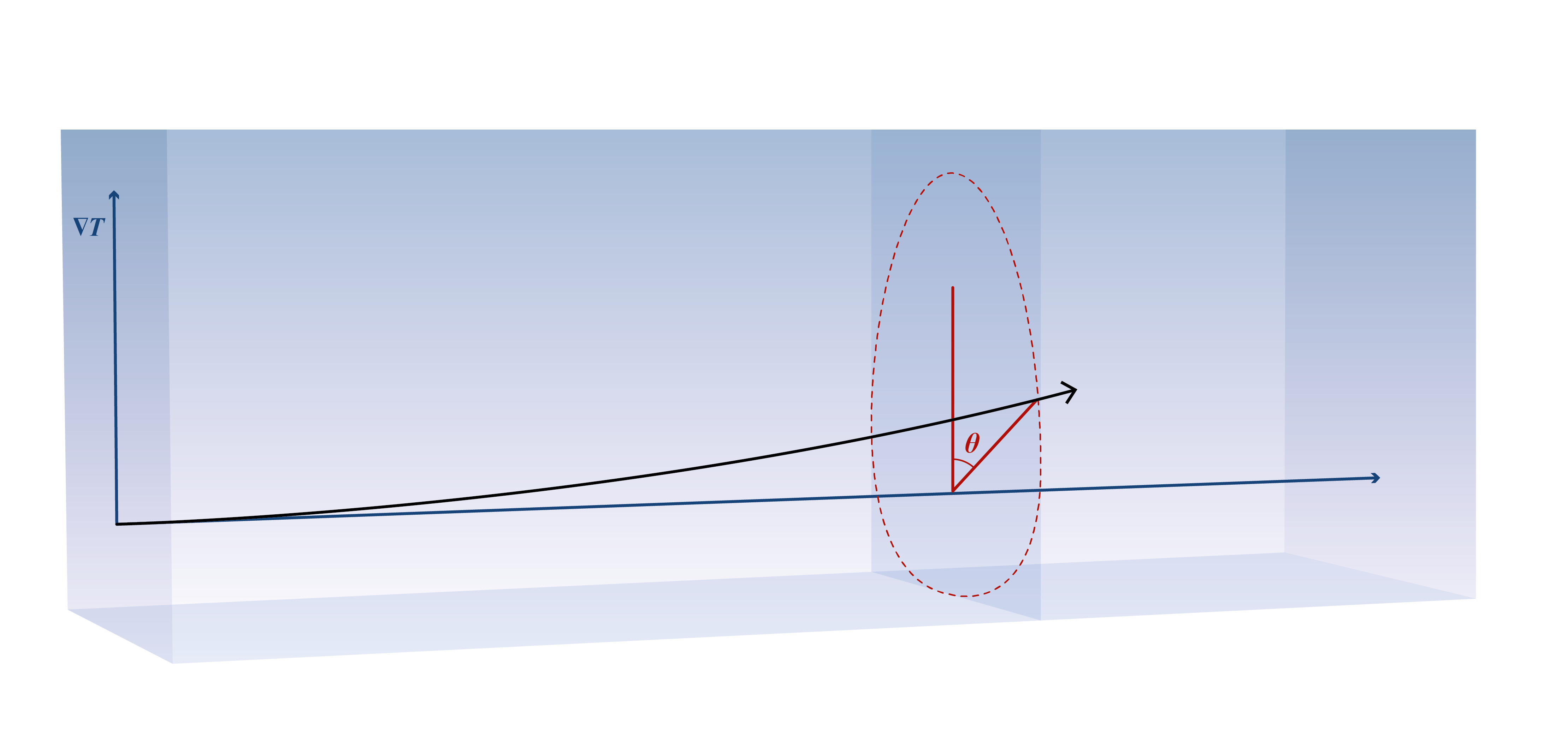}}
    \vspace{-1cm}
    \caption{An illustration of a single parton evolving in the presence of an inhomogeneous static slab of matter. The initial jet direction gets modified due to the presence of the temperature gradient. This modification depends on the angle $\theta$ between $\tvec \na T$ and the transverse momentum accumulated during the evolution, as shown by the dashed shape.}
    \label{fig:nice_fig}
\end{figure}

In this section, we proceed to discuss the properties of the final jet momentum distribution given by \eqref{eq:compare_to_BDMPS} (or equivalently \eqref{dNBDMPSZ}). Heuristically, the leading effect of the matter gradients is that the broadening becomes anisotropic, and the final jet momentum distribution is direction dependent, see Fig.~\ref{fig:nice_fig}. 
In other words, propagating through an inhomogeneous matter, probes pick up an additional transverse momentum, and even on average it is non-zero due to the non-trivial matter structure.
The modifications to the jet structure are thus qualitatively different from the ones observed in the case of a homogeneous background. In this section, we will analyze the properties of the distribution \eqref{eq:compare_to_BDMPS} and illustrate our results with simple numerical estimates for the broadening probability, using phenomenologically relevant parameters.
Let us first notice that the broadening of the jet distribution has to be unitary -- at a fixed energy, the number of jets (partons) cannot be changed and the in-medium propagation results only in a reshuffling of the underlying momentum modes.
To check that, one can consider the value of the $\tvec{x}$-space distribution at $\tvec x = \0$, which corresponds to the volume integral of the momentum-space distribution.
At this point, the $\tvec{x}$-space dipole potential is zero $\mathcal{V}(\tvec 0)=0$ as well as its transverse gradient $\tvec{\na}\mathcal{V}(\tvec 0)=\0$, and the full distribution \eqref{eq:global} is equal to $\int\,d^2\tvec{p}\,\frac{dN^{(0)}}{d^2\tvec{p}\,dE}$. Thus, the unitarity condition is satisfied by the broadened jet momentum distribution even in an inhomogeneous medium
\bea
\int\,d^2\tvec{p}\,\frac{d\N}{d^2\tvec{p}dE}=\frac{d\N}{d^2\tvec{x}dE}\Bigg|_{\x=0}=\frac{d\N^{(0)}}{d^2\tvec{x}dE}\Bigg|_{\x=0}=\int\,d^2\tvec{p}\,\frac{d\N^{(0)}}{d^2\tvec{p}dE}\,.
\eea

One could also notice that in (\ref{dNBDMPSZ}) the initial distribution is not fully factorized from probe-medium interactions. Indeed, in the uniform case, the $\tvec x$-space form of the final jet distribution can be represented as a product of the initial distribution with an interaction factor. The gradient effects in turn enter not only through the probe-medium interactions, but also through a shift of the argument in the initial hard distribution $\frac{dN^{(0)}}{d^2\tvec{p}\,dE}$. Thus, the relation between the final and initial distributions involves an operator
\begin{align}
\label{fact}
\frac{d\N}{d^2\tvec{x}dE}&=\cP\left(\tvec x\right)\hat{\mathsf S }\left(\tvec x\right)\frac{d\N^{(0)}}{d^2\tvec{x}dE}\,,
\end{align}
where 
\begin{align}
\cP(\x)=\exp\left\{-\mathcal{V}\left(\tvec{x}\right)L\right\}\left[1-\frac{iL^3}{6E}\tvec\na\mathcal{V}\left(\tvec{x}\right)\cdot\hat{\tvec{g}}\,\mathcal{V}\left(\tvec x\right)\right]\notag
\end{align}
is a transform of $\cP(\p)$, the (broadening) probability for a particle of energy $E$ to acquire transverse momentum $\p$ due to propagation in the medium for a distance $L$, and 
\begin{align}
\hat{\mathsf S}\left(\tvec x\right)=1+\frac{iL^2}{2E}\,\hat{\tvec{g}}\,\mathcal{V}\left(\tvec x\right)\cdot\tvec\na\simeq \exp\left\{\frac{iL^2}{2E}\,\hat{\tvec{g}}\,\mathcal{V}\left(\tvec x\right)\cdot\tvec\na\right\}\notag
\end{align}
is a shift operator being an identity operator in the absence of gradients. One can immediately see that $\cP(\p)$ is self-normalized, and that $\hat{\mathsf S}\left(\tvec x\right)$ does not change particle number at a given fixed energy.

It is instructive to compare the resummed final state distribution to the one obtained at the first order in opacity by evaluating the leading moments:
\begin{align}
\left\langle F(\tvec p)\right\rangle=\frac{\int\,d^2\tvec p\,F(\tvec p)\,\frac{d\N}{d^2\tvec{p}dE}}{\int\,d^2\tvec p\,\frac{d\N^{(0)}}{d^2\tvec{p}dE}}\,,
\end{align}
where $F(\tvec p)$ is an arbitrary function. Following \cite{Sadofyev:2021ohn}, we focus on a Gaussian initial distribution
\begin{align}
E\frac{d\N^{(0)}}{d^2\tvec{p}\,dE}&=\frac{f(E)}{2\pi w^2}e^{-\frac{p_\perp^2}{2w^2}}\,,
\end{align}
where $f(E)$ is an unspecified energy dependence, and $w$ is the characteristic width. We also assume that the matter is uniform in the longitudinal direction. One generally expects that the terms linear in gradients cannot modify the scalar moments, so $\langle p_\perp^{2k}\rangle = \langle p_\perp^{2k}\rangle_{\na \r=\na \mu^2=\0}$, and we may focus on the odd moments $\langle \p\:p_\perp^{2k}\rangle$ sensitive to the anisotropy\footnote{For an earlier attempt to study the matter anisotropy with directional jet quenching, see \cite{Majumder:2006wi}.}, see e.g. \cite{Sadofyev:2021ohn,Antiporda:2021hpk}.

Starting with the simplest case of the averaged momentum $\left\langle\tvec p\right\rangle$ at the given energy,
we find it convenient to use the opacity expanded form of the distribution (\ref{MR1}) rather than its resummed form. Doing so, one finds that the moment is controlled by the following integral
\begin{align}
\label{averagep}
\frac{1}{2\pi w^2}\int\,d^2\tvec p\,p^\a\,p_m^\b\, e^{-\frac{p_{in\perp}^2}{2w^2}}&=
\d^{\a\b}w^2+\sum_{i=1}^{N}q^\a_i \sum_{j=1}^{m-1}q^\b_j\,,
\end{align}
where $\tvec{p}_n=\tvec{p}_f-\sum\limits_{m=n}^{N} \tvec{q}_m$, $\tvec{p}_{in}=\tvec{p}_1$, and $\a$ and $\b$ are the indices in the transverse 2D space. We also notice that in the absence of the gradients the averaged momentum is zero due to the lack of a preferred direction in the problem, and the corresponding part of (\ref{MR1}) gives zero. Thus, for the contribution at the $N$th order in opacity, we find
\begin{align}
\left\langle p^\a\right\rangle^{(N)}&=\frac{L^{N+1}}{E\,(N+1)!}\prod\displaylimits_{n=1}^N\left[\int \frac{d^2q_n}{(2\pi)^2}\right] 
\sum_{m=1}^N\left[\left(\sum_{i=1}^{N}q^\a_i \sum_{j=1}^{m-1}q^\b_j\right)
\sum_{k=m}^N
\hat{g}^\b_k\right]\notag\\
&\hspace{2cm}\times(-1)^N\mathcal{V}_1(\tvec{q}_{1})\,...\, \mathcal{V}_N(\tvec{q}_{N})\,,
\end{align}
where we have additionally used that $\int\,d^2\tvec{q}_n\,\mathcal{V}_n(\tvec{q}_n)=0$ removing the $w^2$ term. For any $N>1$ this expression is zero due to the same property of $\mathcal{V}_n(\tvec{q}_n)$, since the integrand involves only two powers of $\tvec{q}_i$, and at least one of the dipole potential averages to zero. In turn, the $N=1$ contribution is zero since it involves only the $w^2$ term. Thus, we find that $\left\langle \tvec p\right\rangle=\0$ to all orders in opacity.

The first non-zero moment of the momentum broadening distribution at the first order in opacity is $\left\langle \tvec p\:p_\perp^2\right\rangle$ \cite{Sadofyev:2021ohn}. We can again perform the final momentum averaging, and find
\begin{align}
\label{averagep3}
\frac{1}{2\pi w^2}\int\,d^2\tvec p\,p^\a\,p_m^\b\,p_\perp^2\, e^{-\frac{\tvec{p}_{in}^2}{2w^2}}&=\left[4w^4+w^2\left(\sum_{i=1}^{N}\tvec{q}_i\right)^2\right]\d^{\a\b}+4w^2\left(\sum_{i=1}^{N}q^\a_i\right)\left( \sum_{j=1}^{m-1}q^\b_j\right)\notag\\
&\hspace{-3cm}+2w^2\left(\sum_{i=1}^{N}q^\a_i\right)\left( \sum_{j=1}^{N}q^\b_j\right)+\left(\sum_{i=1}^{N}q^\a_i\right)\left( \sum_{j=1}^{m-1}q^\b_j\right)\left(\sum_{l=1}^{N}\tvec{q}_l\right)^2\,.
\end{align}
Thus, one can see that the only non-zero contributions may come at $N=1$
\begin{align}
\left\langle p^\a\:p_\perp^2\right\rangle^{(1)}&=
-\frac{w^2L^2}{E}\,\hat{g}^\a\int \frac{d^2\tvec q}{(2\pi)^2}\,q_\perp^2\,\mathcal{V}(\tvec{q})\,
\end{align}
and at $N=2$ 
\begin{align}
\left\langle p^\a\:p_\perp^2\right\rangle^{(2)}&=
\frac{L^3}{6E}\int \frac{d^2\tvec{q}_1}{(2\pi)^2}\,q_{1\perp}^2\,\mathcal{V}(\tvec{q}_1)\,\hat{g}^\a\int \frac{d^2\tvec{q}_2}{(2\pi)^2}\,q_{2\perp}^2\,\mathcal{V}(\tvec{q}_2)\,,
\end{align}
while the higher orders in opacity decouple. In the case of the GW potential, the terms proportional to $\tvec\na\r$ are divergent, and should be regularized at some scale. Comparing with \cite{Sadofyev:2021ohn}, we set $|\tvec{q}_{\text{max}}|=\sqrt{E\m}$, and find
\begin{align}\label{eq:p3_GLV}
\left\langle p^\a\:p_\perp^2\right\rangle&=\frac{w^2L^2\m^2}{E\,\l}\frac{\na^\a\r}{\r} \ln\frac{E}{\m}+\frac{L^3\m^4}{6E\,\l^2}\frac{\na^\a\r}{\r}\left(\ln\frac{E}{\m}\right)^2\,,
\end{align}
where $\l=\frac{1}{\r\s_0}$ with $\s_0=\int\frac{d^2\tvec{q}}{(2\pi)^2}\,\mathcal{C}\,|v(q_\perp^2)|^2$. The first term in the expression above precisely agrees with the result obtained in \cite{Sadofyev:2021ohn}. In turn, the second term is new and indicates that the higher $N$ contributions to the transverse momentum moments with integer $k\geq1$ are generally dominating as long as the potential integrals are divergent. One should also notice that the second term is independent of $w$, and can be identified as a purely final state effect (described by the broadening probability $\cP(\p)$), since in the limit $w\to0$ the initial distribution is constant in coordinate space and not affected by the shift operator $\hat{\mathsf S}$. In general, non-zero odd moments are the primary effect of the hydrodynamic gradients on the jet broadening, and they can be used to learn about the medium profile with jets. We leave the moments with general real $k$ and their phenomenological implications for future studies. 

We now turn back to the partial factorization, and focus on the final state effects. The corresponding portion of (\ref{fact}) is the self-normalized broadening probability $\cP(\tvec{p})$, which has multiple phenomenological applications \cite{Caucal:2019uvr,Blaizot:2013vha,Kutak:2018dim,Barata:2021byj,Soudi:2021aar}. Its Fourier transform can be obtained from (\ref{dNBDMPSZ}) if we set $E\frac{d\N^{(0)}}{d^2\tvec{x}dE}$ to be a constant in the transverse directions, decoupling in this way the shift operator $\hat{\mathsf S}(\tvec x)$:
\begin{align}
\label{Prob}
\cP(\tvec{p})&=\int\,d^2\tvec{x}\,e^{-i\tvec p\cdot\tvec x}e^{-\mathcal{V}\left(\tvec{x}\right)L}\left[1-\frac{iL^3}{6E}\tvec\na\mathcal{V}\left(\tvec{x}\right)\cdot\hat{\tvec{g}}\mathcal{V}\left(\tvec x\right)\right]\,.
\end{align}
At this point one has to specify the particular model for the medium, fixing the dipole potential, and evaluate $\cP(\tvec p)$ explicitly. We consider the GW model, and in this case the potential reads
\begin{equation}\label{eq:GW_v}
\mathcal{V}^{\rm GW}(x_\perp) = \frac{\mathcal{C}g^4\r}{4\pi\mu^2}\left(1-\m\, x_\perp K_1(\mu\, x_\perp)\right) \, ,
\end{equation}
However, even in this simple case and in the absence of gradients, the Fourier transform cannot be written in a closed form, restricting phenomenological applications. 
By this reason, it is instructive to consider $\cP(\tvec p)$ in some limiting regimes. 
If the gradients are zero, then the probability (\ref{Prob}) is a function of two dimensionless combinations of the parameters and momentum: $\frac{p_\perp}{\m}$ and $\chi\equiv\frac{\mathcal{C}g^4\r}{4\pi\m^2}L$, where $\chi$ is the opacity in the GW model. The dipole potential $\mathcal V(x_\perp)$ tends to a constant at large $\m\,x_\perp$, and the large $x_\perp$ limit of the integration is controlled by the fast oscillation of the Fourier exponential. For sufficiently large momenta such that $p_\perp\gg\m$, one can expand the dipole potential in powers of $\m\,x_\perp$, and for the GW model it reads
\begin{align}
\label{Vexp}
\frac{4L}{\chi}\mathcal{V}^{\rm GW}(x_\perp) = \m^2x_\perp^2\log\frac{4e^{1-2\g_E}}{\m^2x_\perp^2}+\mathcal{O}\left(\m^4x^4_\perp
\right) \, ,
\end{align}
where $\g_E$ is the Euler constant. One should notice that the first term in the expansion is common between most of the medium models, since it is fixed by the ultraviolet (UV) Coulomb-like behavior 
for large momentum exchanges, see e.g. \cite{Barata:2020sav,Barata:2020rdn,Djordjevic:2009cr}.  

If we further require that $p_\perp^2\gg\chi\,\m^2$, then the exponential of the dipole potential itself can be expanded,  and, after expanding the in-medium potentials, we find
\begin{align}
\label{Phard}
\cP(\tvec{p})&\simeq\int\,d^2\tvec{x}\,e^{-i\tvec p\cdot\tvec x}\Bigg[1-\mathcal{V}\left(x_\perp\right)L+\frac{1}{2}\mathcal{V}^2\left(x_\perp\right)L^2-\frac{iL^3}{6E}\tvec\na\mathcal{V}\left(x_\perp\right)\cdot\hat{\tvec{g}}\mathcal{V}\left(x_\perp\right)\Bigg]\notag\\
&\hspace{0cm}\simeq\frac{4\pi\m^2\chi}{p_\perp^4}+\frac{16\pi\mu^4\chi^2}{p_\perp^6}\left(\log\frac{p_\perp^2}{\m^2}-2\right)\notag\\
&\hspace{4cm}+\frac{4\pi \m^4 \chi^2 L}{3E}\left[\frac{\tvec\na\r}{\r}\left(\log\frac{p_\perp^4}{\mu^4}-4\right)-\frac{\tvec\na\m^2}{\m^2}\right]\cdot\frac{\tvec{p}}{p_\perp^6}\, ,
\end{align}
where we have omitted terms proportional to $\d^{(2)}(\tvec p)$ since the argument is away from zero. The first two terms in this expansion come from the homogeneous case, with the first giving the dominant Coulomb tail expected at large momentum transfers. In turn, the gradient term has an odd (negative) power 
of $p_\perp$ and can only be generated in an inhomogeneous medium. In fact, one can compare the asymptotic structure of \eqref{Phard} directly with the first non-trivial moment 
\eqref{eq:p3_GLV} in the limit of infinitely narrow source (i.e. $w\to 0$). The only contribution to \eqref{eq:p3_GLV} in this limit comes from the $N=2$ term, which gets a double Coulomb logarithmic enhancement. Indeed, looking at the last term in \eqref{Phard}, one may notice that its dominant contribution to $\langle \p \:  p_\perp^2 \rangle$ scales as $\left|\frac{\tvec \na \r}{E\r}\right| \int dp_\perp \frac{\log\,p_\perp}{p_\perp}$, and with a similar UV regularization it gives $\left|\frac{\tvec \na \r}{E\r}\right|  \left(\log \frac{E}{\mu}\right)^2$ in qualitative agreement with the previous result.

Expanding the probability distribution $\cP$ in two parameters, one has to pay a particular attention to the applicability of the result. Indeed, if the large momentum expansion parameter $\frac{\m}{p_\perp}$ is too small, the gradient contributions could take the leading role. Here we assume that this is not the case, and that the first order gradient correction appearing in the last term of \eqref{Phard} is at least smaller than the dominant Coulomb tail contribution, although it can compete with the second term in the expression.

Another kinematic regime commonly considered in the literature \cite{Barata:2020rdn,Caucal:2019uvr} corresponds to the intermediate momentum region $\m^2\ll p_\perp^2\leq\chi\,\m^2$. In this case, the exponential of the dipole potential cannot be expanded, while the approximation (\ref{Vexp}) can still be utilized. Moreover, the weak logarithmic dependence of the expanded dipole potential on $\tvec x$ can be neglected. Indeed, one can introduce a new scale $\mathcal{Q}^2\gg\m^2$ such that
\begin{align}
\label{Vexp01}
\frac{4L}{\chi}\mathcal{V}^{\rm GW}(x_\perp) \simeq \m^2x_\perp^2\left(\log\frac{\mathcal{Q}^2}{\m^2}+\log\frac{4e^{1-2\g_E}}{\mathcal{Q}^2x_\perp^2}\right)\simeq\m^2x_\perp^2\log\frac{\mathcal{Q}^2}{\m^2}\,.
\end{align}
In practice, it can be fixed up to an overall coefficient based on phenomenological arguments, see e.g.~\cite{Barata:2020rdn,Barata:2021wuf} for further discussion. In this regime, the dipole potential is quadratic, and the leading broadening probability (\ref{Prob}) becomes Gaussian. This is a regime of multiple soft interactions, which is widely used in phenomenological models for jet quenching, see e.g. \cite{Caucal:2019uvr,Casalderrey-Solana:2014bpa, Blaizot:2013vha}, and often referred to as BDMPS-Z/ASW approximation \cite{Zakharov:1997uu,Armesto:2003jh}. Taking into account the gradient effects, we find
\begin{align}
\label{Psoft}
\cP(\tvec{p})&\simeq\int\,d^2\tvec{x}\,e^{-i\tvec p\cdot\tvec x}e^{-\frac{1}{4}\chi\m^2x_\perp^2\log\frac{\mathcal{Q}^2}{\m^2}}\left[1-\frac{i\m^4\chi^2L}{48E}\log\frac{\mathcal{Q}^2}{\m^2}\,x_\perp^2\,\tvec{x}\cdot\left(\frac{\tvec\na\r}{\r}\log\frac{\mathcal{Q}^2}{\m^2}-\frac{\tvec\na\m^2}{\m^2}\right)\right]\notag\\
&\hspace{-1cm}=\frac{4\pi}{\chi\m^2\log\frac{\mathcal{Q}^2}{\m^2}}\Bigg[1+\frac{L}{6E}\,\frac{p_\perp^2-2\chi\m^2\log\frac{\mathcal{Q}^2}{\m^2}}{\chi\m^2\log\frac{\mathcal{Q}^2}{\m^2}}\,\left(\frac{\tvec \na\r}{\r}-\frac{1}{\log\frac{\mathcal{Q}^2}{\m^2}}\frac{\tvec \na\m^2}{\m^2}\right)\cdot\tvec{p}\Bigg]e^{-\frac{p_\perp^2}{\chi\m^2\log\frac{\mathcal{Q}^2}{\m^2}}}\,.
\end{align}
The usual Gaussian result gets an overall $\p$-dependent modulation by a new term containing all the gradient effects. We note that the effect of this contribution should be more important when $p^2_\perp\sim \chi \mu^2 \log\frac{\mathcal{Q}^2}{\m^2} $, roughly at the peak of the $p_\perp$-distribution~\cite{Barata:2020rdn}.
In this region, the leading behavior of $\cP$ is fully described neither by a single hard scattering nor by the multiple soft scatterings
, but rather there is a competition between these two regimes. Thus, one may expect a more pronounced deviation from the homogeneous solution in this region. We numerically verify this observation below in Fig.~\ref{fig:P_dist}.

Finally, for $\m^2\ll p_\perp^2\leq\chi\,\m^2$, we can also consider the small logarithmic correction to the dipole potential, neglected in (\ref{Vexp01}), as a perturbation
\begin{align}
\label{Vexp02}
\mathcal{V}^{\rm GW}(x_\perp) \simeq \frac{\chi\m^2}{4L}x_\perp^2\left(\log\frac{\mathcal{Q}^2}{\m^2}+\log\frac{4e^{1-2\g_E}}{\mathcal{Q}^2x_\perp^2}\right)=\mathcal{V}_0(x_\perp)+\d\mathcal{V}(x_\perp)\,.
\end{align}
This approach to capture the leading effects beyond the quadratic Gaussian approximation is known as the improved opacity expansion (IOE), see e.g. \cite{Mehtar-Tani:2019ygg,Mehtar-Tani:2019tvy,Barata:2020sav}. Substituting the potential into the broadening probability, we write it as an expansion
\begin{align}
\label{ProbIOE}
\cP_{\rm IOE}(\tvec{p})&=\cP_{0+1}(\tvec{p})+\cP_{1+0}(\tvec{p})+\cP_{2+0}(\tvec{p})+\cP_{1+1}(\tvec{p})+\cP_{2+1}(\tvec{p})+\cP_{3+1}(\tvec{p})+...\,,
\end{align}
where the first number in the subscript corresponds to the order in $\d\mathcal{V}$, and the second one counts the order in gradients. The leading contribution $\cP_{0+1}(\tvec{p})$ is given by (\ref{Psoft}), the term $\cP_{1+0}(\tvec{p})$ has been discussed in details in \cite{Barata:2020rdn}, and we have to evaluate the mixed contributions. The full derivation of the sub-leading terms goes beyond the scope of this paper, and we focus on their large momentum regime. That will allow us to see that the terms sub-leading in $\d\mathcal{V}$ give the correct large momentum limit (\ref{Phard}) even in the presence of gradients, {\it c.f.} \cite{Barata:2020rdn}. 

All the $\tvec{x}$-integrals entering (\ref{ProbIOE}) up to the second order can be obtained from three master integrals if we act on them with the appropriate number of momentum derivatives. They read
\begin{align}
I_1(\tvec{p},a)&=\int\,d^2\tvec{x}\,e^{-i\tvec p\cdot\tvec x}e^{-ax_\perp^2}=\frac{\pi}{a}e^{-\frac{p_\perp^2}{4a}}\, ,\notag\\
I_2(\tvec{p},a,b)&=\int\,d^2\tvec{x}\,e^{-i\tvec p\cdot\tvec x}e^{-ax_\perp^2}\,\log\frac{1}{b^2x_\perp^2}=\frac{\pi}{a}e^{-\frac{p_\perp^2}{4a}}\left[Ei\left(\frac{p_\perp^2}{4a}\right)-\log\frac{b^2p_\perp^2}{4a^2}\right]\notag\\
&\simeq\frac{4\pi}{p_\perp^2}+\frac{16\pi a}{p_\perp^4}+\mathcal{O}\left(\frac{a^2}{p_\perp^6}\right)\, ,\notag\\ I_3(\tvec{p},a,b)&=\int\,d^2\tvec{x}\,e^{-i\tvec p\cdot\tvec x}e^{-a x_\perp^2}\,\log^2\frac{1}{b^2 _\perp x_\perp^2}
\notag\\
&\simeq\frac{16\pi}{p_\perp^2}\left(\g_E+\log\frac{p_\perp}{2b}\right)+\frac{64\pi a}{p_\perp^4}\left(\g_E-1+\log\frac{p_\perp}{2b}\right)+\mathcal{O}\left(\frac{a^2}{p_\perp^6}\log\frac{p_\perp}{2b}\right)\,,
\end{align}
where we omit the explicit form of $I_3$ for brevity. In the large $p_\perp$ limit, the momentum derivatives acting on a term reduce its contribution removing powers of $p_\perp$. Consequently, one can compare the relative importance of the different contributions in (\ref{ProbIOE}) by counting the powers of $\tvec{x}$ entering through $\mathcal{V}$ and $\d\mathcal{V}$ outside the Gaussian exponential. However, since the structure of the gradient terms in (\ref{Prob}) involves gradients of the dipole potential, we have to consider higher order terms in $\d\mathcal{V}$ to be sure that we keep all the relevant contributions. 

We first notice that the leading term in the IOE expansion $\cP_{0+1}(\tvec{p})$ is exponentially suppressed for $p_\perp^2\gg\chi\m^2$, while $\cP_{1+0}(\tvec{p})\propto\frac{1}{p_\perp^4}$ gives the leading Coulomb term in this limit. Similarly, the sub-leading non-gradient contribution satisfies $\cP_{2+0}(\tvec{p})\propto\frac{1}{p_\perp^6}\log p_\perp$. 
The mixed gradient terms in turn scale as $\cP_{1+1}(\tvec{p})\propto\frac{1}{p_\perp^5}$ and $\cP_{2+1}(\tvec{p})\propto\frac{1}{p_\perp^5}\log p_\perp$, while one can check that the next contribution $\cP_{3+1}(\tvec{p})\propto\frac{1}{p_\perp^6}\log^2p_\perp$ is suppressed both by a higher power of inverse momentum as well as by the gradients, and can be omitted. Finally, we find
\begin{align}
\label{PIOE}
\cP_{\rm IOE}(\tvec{p})&\simeq\frac{4\pi\m^2\chi}{p_\perp^4}+\frac{16\pi\mu^4\chi^2}{p_\perp^6}\left(\log\frac{p_\perp^2}{\m^2}-2\right)\notag\\
&\hspace{1cm}+\frac{4\pi \m^4 \chi^2 L}{3E}\left[\frac{\tvec\na\r}{\r}\left(\log\frac{p_\perp^4}{\mu^4}-4\right)-\frac{\tvec\na\m^2}{\m^2}\right]\cdot\frac{\tvec{p}}{p_\perp^6}\,,
\end{align}
which precisely agrees with the large momentum limit (\ref{Phard}), showing that the IOE covers both limits considered above. It could be also noticed that the logarithmic terms in $\cP_{\rm IOE}(\tvec{p})$ are sensitive to a cancellation of $\mathcal{Q}$ between terms of different orders in the IOE, {\it c.f.}~\cite{Barata:2021wuf}.

Finally, to have a more quantitative understanding of the broadening distribution $\cP(\p)$, we evaluate \eqref{Prob} using the full GW potential given in \eqref{eq:GW_v}. We also assume that the nuclear matter is near to equilibrium, and its properties are controlled by a single parameter.
For instance, in the case of QGP in the large temperature limit, one can use parametric scaling motivated by the equilibrium thermodynamics $\mu\propto g T$ and $\rho \propto T^3$, then
\begin{align}
\label{scaling}
\frac{\tvec \na \r}{\r} = 3  \frac{\tvec\na T}{T}  \, , \quad  \frac{\tvec \na \mu^2}{\mu^2} = 2  \frac{\tvec \na T}{T} \, .
\end{align}
The probability $\cP(\p)$ is now a function of $\theta$, the angle between $\p$ and $\tvec \na T$, and a dimensionless quantity $c_T\equiv \left|\frac{\tvec \na T}{ET}\right|\ll 1$, see Fig.~\ref{fig:nice_fig}.  
Performing the remaining angular integration, we can write $\cP$ as
\begin{align}\label{eq:num_fin_sub0}
\cP(\p)&=    2\pi \int\limits^\infty_0 dx_\perp \, x_\perp\, e^{-\mathcal{V}^{\rm GW}(x_\perp) L} \Bigg\{J_0(p_\perp x_\perp)-\frac{\chi^2 \mu^2 L}{6}\,c_T\,x_\perp\,K_0(\m\,x_\perp)\,J_1(p_\perp x_\perp)\notag\\
&\hspace{1cm}\times\left[1-3\mu\,x_\perp K_1(\mu\,x_\perp)+\mu^2 x_\perp^2K_2(\mu\,x_\perp) \right] \cos{\theta} \Bigg\} \, .
\end{align}
For the particular case of the GW model, $\mathcal{V}^{\rm GW}(x_\perp)$ asymptotically tends to a constant. Therefore, the integral in \eqref{eq:num_fin_sub0} needs to be regulated for values of $x_\perp\gg \frac{1}{\mu}$. In particular, one should remove the non-scattering probability term associated with the asymptotic behavior of $\mathcal{V}^{\rm GW}(x_\perp)$, so that only genuine broadening contributions are taken into account. Thus, following e.g.~\cite{Feal:2018jbm,Barata:2020rdn}, we consider a regularized broadening distribution
\begin{align}\label{eq:num_fin}
\cP_r(\p)&=  \cP(\p) -   2\pi \int\limits^\infty_0 dx_\perp \, x_\perp\, e^{-\mathcal{V}^{\rm GW}(\infty) L} \, J_0(p_\perp x_\perp)\, ,
\end{align}
which differs from \eqref{eq:num_fin_sub0} by a singular term (delta function) at $\p=\0$.

\begin{figure}[h]
\hspace*{-0.75cm}  
    \centerline{
    \includegraphics[width=1\textwidth]{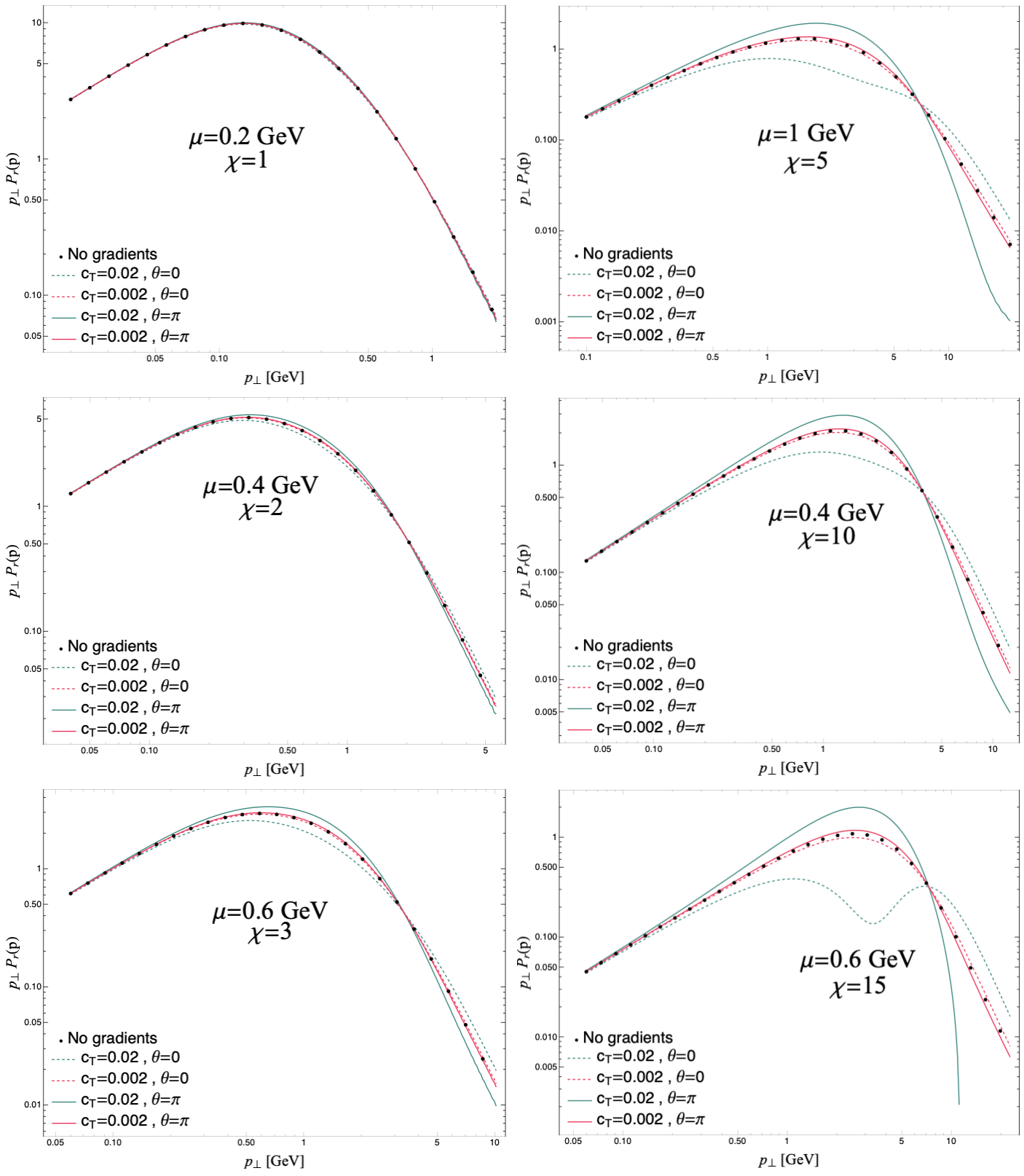}}
    \vspace{-0.5cm}
    \caption{Numerical evaluation of \eqref{eq:num_fin} for six different medium parametrizations. Each setup is characterized by its values for the opacity $\chi$, the Debye mass $\mu$, and a medium length $L=5 \, {\rm fm}$. The values considered for the medium parameters are inspired by the ones typically found in the literature~\cite{Vitev:2008vk, Xu:2014ica, Sievert:2019cwq, Feal:2019xfl, Barata:2020rdn, Antiporda:2021hpk}.}
    \label{fig:P_dist}
\end{figure}

In Fig.~\ref{fig:P_dist} we present numerical evaluations of \eqref{eq:num_fin} for several sets of parameters inspired by the particular values used for phenomenological estimates in the literature, see e.g. \cite{Vitev:2008vk, Xu:2014ica, Sievert:2019cwq, Feal:2019xfl, Barata:2020rdn, Andres:2020vxs, Antiporda:2021hpk}. 
However, we have to additionally stress here that \eqref{eq:num_fin} has been evaluated in the simplest limit of a longitudinally uniform slab of nuclear matter with the medium parameters varying in the transverse directions. Thus, any realistic phenomenological applications would require further generalizations of this broadening distribution. Each plot is characterized by a value for the medium opacity $\chi$, the medium length size $L$, and the Debye mass $\mu$. Gradient effects are controlled by $\theta$ and $c_T$. We take the medium size to be $L=5 \, {\rm fm}$, the typical size of nuclear matter, for all the plots. To explore the sensitivity of $\cP_r(\p)$ to the matter gradients, we chose to either take $c_T=0.02$ or $c_T=0.002$ with the temperature gradient either parallel ($\theta=0$) or antiparallel ($\theta=\pi$) to $\tvec{p}$. The particular values of $c_T$ 
should be understood in the following way. In the hydrodynamic regime the temperature gradient is expected not to be large, i.e. $|\tvec{\na}T|<T^2$, while the characteristic jet energy $E$ should be much larger than the temperature $T$. If one takes $\left|\frac{\tvec \na T}{T^2}\right|\sim 1$ and $\left|\frac{\tvec \na T}{T^2}\right|\sim 0.1$, then the $c_T$ values correspond to $E\sim 50 \, T$, and one can further scale the value of the gradient and jet energy together.
Finally, when numerically evaluating \eqref{eq:num_fin}, we perform the remaining integration up to a scale of the order of $\frac{1}{\mu}$. This ultraviolet cutoff was systematically varied in order to ensure that the final results do not depend on its particular choice.

At the qualitative level, we observe that the gradient effects are small 
for lower opacity,
since the gradient corrections in \eqref{eq:num_fin} are proportional to $\chi^2$.  
Comparing all the results shown, we also confirm the claim made above that gradient effects should become more important around the peak of the distribution. Indeed, for the higher opacity scenarios, gradient effects lead to a sizable suppression of the broadening peak (for $c_T=0.02$ and $\theta=0$), resulting in an increase of the broadening probability at higher $p_\perp$. Conversely, in the case where $\p$ is antiparallel to $\tvec \na T$, we observe an increase of the distribution peak, leading to a depletion of higher momentum modes. Another feature shared by all the setups is the insensitivity of the result to gradient effects at small $p_\perp$, due to the fact that the inhomogeneous term always couples to an odd power of $\p$. Finally, it should be also mentioned that the large momentum tail of the distribution is not modified by the gradient corrections, as seen from (\ref{PIOE}). This is indeed true for the numerically evaluated distributions, but the corresponding large $p_\perp$ region is not presented in Fig.~\ref{fig:P_dist}.

\section{Conclusions and Outlook}\label{ref:conclusion}
In this work, we have derived the parton momentum broadening distribution in an inhomogeneous dense nuclear matter. The final distribution has been obtained using both the GLV and BDMPS-Z formalisms. In the homogeneous case, these two approaches considered above were proven to agree order by order in opacity, see e.g. \cite{Wiedemann:2000ez, Wiedemann:2000za, Gyulassy:2002yv, Qiu:2003pm}. Here we make the next step, showing that the agreement holds even if the matter is not translationally-invariant in the transverse directions. Importantly, the effects of the medium inhomogeneity only emerge once the medium averaging is performed. As a consequence, all the results for the jet-medium interactions which are independent of the averaging procedure stay unmodified with respect to the homogeneous case.
Qualitatively, the modification of the resulting broadening distribution (comparing to the homogeneous baseline) becomes relevant around and above its characteristic peak, introducing a novel angular dependence. As a consequence, the gradient effects can have sizable impact in phenomenological applications, even for momentum averaged quantities.

Our results for the jet broadening can be extended to the case of inelastic energy loss in a dense inhomogeneous medium. Primarily, such an exercise will allow to study how the gradient effects alter the in-medium gluon production rate, giving rise to a non-trivial angular structure of the radiation. One may expect that the effective time factorization of the emission spectrum found in the limit of soft induced radiation in homogeneous media~\cite{Mehtar-Tani:2012mfa,Blaizot:2013vha} would be modified in accordance with~\eqref{fact}. Understanding the resulting new factorized form is critical for some aspects of jet quenching phenomenology~\cite{Blaizot:2013vha,Caucal:2019uvr,Casalderrey-Solana:2014bpa}.

Another important goal we leave for the future studies is to design a jet observable sensitive to the matter effects discussed in this manuscript, {\it c.f.}~\cite{Antiporda:2021hpk}. Particularly, it would be interesting to explore 
to what extent the jet substructure grooming/tagging techniques, see e.g.~\cite{Larkoski:2014wba, Chien:2016led,Larkoski:2017jix, Apolinario:2017qay, Caucal:2021cfb,Karlberg:2021kwr} could be used to look for matter inhomogeneity effects on the angular spectrum of the hardest substructures within the jet. Such new observables would provide a window to further probe the medium properties locally along the ideas of jet tomography.

We also note that the presented results are independent of the physical process which generates the background medium and, thus, they are in principle applicable to either the QGP phase or the glasma phase in the HIC context~\cite{Ipp:2020mjc,Carrington:2021dvw}. The latter one could be considerably anisotropic, resulting in stronger directional effects. It would be interesting to better understand the relation 
between the jet quenching formalisms considered here and other approaches which can describe jet evolution in inhomogeneous and evolving media, see e.g.~\cite{Agostini:2019avp, Hauksson:2021okc,Altinoluk:2021lvu}. In turn, extension to the DIS context would require the present calculation to be updated to the corresponding kinematical regime with particular focus on the factorization of the jet production and its propagation through the medium, see e.g.~\cite{Armesto:2013fca,Wang:2001ifa, Kang:2013raa, Kang:2014ela, Sirimanna:2021sqx} and references therein.

Finally, it should be mentioned that the dependence of the jet-medium interaction on hydrodynamic gradients could be used to probe the strength of the interactions in the underlying theory. Indeed, comparing the characteristic transport properties and the general jet behavior in an inhomogeneous matter between pQCD and holographic models for jet-medium interactions, see e.g. \cite{Casalderrey-Solana:2014bpa, Casalderrey-Solana:2015vaa, Casalderrey-Solana:2016jvj,Rajagopal:2016uip, Brewer:2017dwd, Brewer:2017fqy, Brewer:2018mpk}, one may hope to identify a set of new observables distinguishing the two regimes. We leave this intriguing opportunity for future work.

\section*{Acknowledgements}
 The authors would like to thank C. Andres, N. Armesto, F. Dominguez, X. Mayo, M. Sievert, I. Vitev, and B. Wu for discussions and comments on this work. This work is supported by European Research Council project ERC-2018-ADG-835105 YoctoLHC; by Maria de Maetzu excellence program under project MDM-2016-0692 and CEX2020-001035-M; by Spanish Research State Agency under project PID2020-119632GB-I00; and by Xunta de Galicia (Centro singular de investigación de Galicia accreditation 2019-2022), by European Union ERDF. The work of A.S. is also supported by the Marie Sklodowska-Curie Individual Fellowship under JetT project (project reference 101032858);

\appendix

\section{The Gradient Corrections at the $N$th Order in Opacity}
\label{App1}

In this Appendix, we present the jet broadening distribution in an inhomogeneous matter at leading orders in opacity, and discuss the general structure of the opacity series. While $N=0$ case is trivial, the leading gradient corrections at $N=1$ have been obtained in \cite{Sadofyev:2021ohn}, and we only quote the result here
\begin{align}
\left\langle \left| M\right|^2 \right\rangle^{(1)} & = -\int\displaylimits_0^{L}\, dz\int \frac{d^2\tvec{q}}{(2\pi)^2}
 \left\{1+ \left[\frac{(\tvec{p}-\tvec{q})}{E}z\right]\cdot\hat{\tvec{g}}\right\}\mathcal{V}_1(\tvec{q})|J(E, \tvec{p}-\tvec{q})|^2\,,
\end{align}
writing it in the notations of this work. 

Turning to the $N=2$ case, we notice that there are five types of contributions, which we have to consider separately. We start with the simplest diagram involving no contact interactions, and the corresponding amplitude reads
\begin{align}
iM_{11} &=\sum_{i_1,i_2}\int \frac{d^2 \tvec{q}_1\,d^2 \tvec{q}_2}{(2\pi)^4}\, \left(it^a_\mathrm{proj} t_{i_1}^a\right)\left(it^b_\mathrm{proj} t_{i_2}^b\right)\,\theta_{2,1}\theta_{1,0}\, 
    e^{-i \tvec{q}_{1} \cdot \tvec{x}_{i_1}}e^{-i \tvec{q}_{2} \cdot \tvec{x}_{i_2}}v_{i_2} (q_2) v_{i_1} (q_1 )
    \notag\\
    &\hspace{1cm}\times e^{-i\frac{(p-q_1-q_2)_\perp^2-p_\perp^2}{2E}z_{i_1}}e^{-i\frac{(p-q_2)_\perp^2-p_\perp^2}{2E}\left(z_{i_2}-z_{i_1}\right)}\,J\left(E,\tvec{p}-\tvec{q}_1-\tvec{q}_2\right) \,,
\end{align}
where the subscript of $M$ indicates that there are two SB insertions. This amplitude has the same form as in the well known homogeneous case, and we omit its derivation. Squaring this contribution and averaging over the quantum numbers, we find
\begin{align}
\left\langle \left| M\right|^2 \right\rangle^{(2)}_{11,11} & =\mathcal{C}^2\int\limits^L_0\,dz_2\int\limits^{z_2}_0\,dz_1\,\int \frac{d^2 \tvec{q}_1\,d^2 \tvec{q}_2}{(2\pi)^4} \Bigg\{1+\frac{\tvec{p}-\tvec{q}_1-\tvec{q}_2}{E}z_1\cdot \hat{\tvec{g}}_1+\frac{\tvec{p}-\tvec{q}_1-\tvec{q}_2}{E}z_1\cdot \hat{\tvec{g}}_2\notag\\
&+\frac{\tvec{p}-\tvec{q}_2}{E}(z_2-z_1)\cdot\hat{\tvec{g}}_2\Bigg\}\,\r_2\,\r_1\,v_2^2(q_{2\perp}^2)v_1^2(q_{1\perp}^2)|J(E, \tvec{p}-\tvec{q}_1-\tvec{q}_2)|^2 
\end{align}
where the $\tvec{x}$-dependence has been transformed to momentum space and we assumed that the matter is longitudinally uniform. Notice that the momentum derivatives of the form $\frac{\pa}{\pa(q-\overline{q})_\a}$, where $\overline{q}$ is the corresponding momentum in the conjugated amplitude, can act only on the LPM phases at the first order in gradients, since the rest of the integrand is symmetric under exchange of the two momenta in each pair. 

Now we turn to the two mixed contributions in the amplitude squared, which involve both SB and DB interactions. It is natural to consider them partially averaged (we choose the contact interaction pair and perform color averaging), then
\begin{align}
iM_{12} &=\mathcal{C}\sum_{i_1,i_2}\int \frac{d^2 \tvec{q}_1\,d^2 \tvec{q}_2\,d^2 \tvec{q}_3}{(2\pi)^6}\, \left(t^a_\mathrm{proj} t_{i_1}^a\right)\,\theta_{2,1}\theta_{1,0}\, 
    e^{-i \tvec{q}_{1} \cdot \tvec{x}_{i_1}}e^{-i \left(\tvec{q}_{2}+\tvec{q}_{3}\right) \cdot \tvec{x}_{i_2}} v_{i_1} (\tvec{q}_1) \mathcal{I}^{12}_{i_2}\left(\tvec{q}_2,\tvec{q}_3\right) 
    \notag\\
    &\times e^{-i\frac{(p-q_1-q_2-q_3)_\perp^2-p_\perp^2}{2E}z_{i_1}}e^{-i\frac{(p-q_2-q_3)_\perp^2-p_\perp^2}{2E}\left(z_{i_2}-z_{i_1}\right)}\,J\left(E,\tvec{p}-\tvec{q}_1-\tvec{q}_2-\tvec{q}_3\right) \,,
\end{align}
where the subscript indicates that the first and second insertions are SB and DB correspondingly, and
\begin{align}
\mathcal{I}^{12}_{i_2}&=2E\int\frac{dq_{3,z}}{(2\pi)}\frac{v_{i_2}\left(\tvec{q}_2,\frac{(p-q_2-q_3)_\perp^2-p_\perp^2}{2E}-q_{3,z}\right)v_{i_2}\left(\tvec{q}_3,q_{3,z}\right)}{\left(p-q_3\right)^2+i\e}\notag\\
&\hspace{1cm}\simeq-\frac{i}{2}v_{i_2}\left(q_{2\perp}^2\right)v_{i_2}\left(q_{3\perp}^2\right)\,.
\end{align}
At $N=2$ this diagram contributes to the amplitude squared multiplied by the diagram with a single SB insertion. Performing the residual averaging, one finds
\begin{align}
\left\langle M_{12}M^*_1 \right\rangle + c.c. &=-\mathcal{C}^2\int\limits^L_0\,dz_2\int\limits^{z_2}_0\,dz_1\,\int \frac{d^2 \tvec{q}_1\,d^2 \tvec{q}_2}{(2\pi)^4} \Bigg\{1+\frac{\tvec{p}-\tvec{q}_1}{E}z_1\cdot \hat{\tvec{g}}_1+\frac{\tvec{p}-\tvec{q}_1}{E}z_1\cdot \hat{\tvec{g}}_2\notag\\
&\hspace{0cm}+\frac{\tvec{p}}{E}(z_2-z_1)\cdot\hat{\tvec{g}}_2\Bigg\}\,\r_2\,\r_1\,v_2^2(q_{2\perp}^2)v_1^2(q_{1\perp}^2)|J(E, \tvec{p}-\tvec{q}_1)|^2\,.
\end{align}
where we have used that the momentum derivatives cancel between the complex conjugated contributions, unless they act on the LMP phases.

Similarly, we can consider the opposite order of the SB and DB insertions, and the corresponding amplitude reads
\begin{align}
iM_{21} &=i\mathcal{C}\sum_{i_1,i_3}\int \frac{d^2 \tvec{q}_1\,d^2 \tvec{q}_2\,d^2 \tvec{q}_3\,dq_{3,z}}{(2\pi)^7}\, \left(t^a_\mathrm{proj} t_{i_3}^a\right)\,\theta_{1,0}\,e^{-i (\tvec{q}_{1}+\tvec{q}_{2}) \cdot \tvec{x}_{i_1}}e^{-i \tvec{q}_{3} \cdot \tvec{x}_{i_3}}  \mathcal{I}^{21}_{i_1}\left(\tvec{q}_1,\tvec{q}_2,\tvec{q}_3,q_{3,z}\right) 
    \notag\\
    &\times e^{-i\frac{(p-q_1-q_2-q_3)_\perp^2-p_\perp^2}{2E}z_{i_1}}\frac{2E\,v_{i_3} (q_3)e^{-iq_{3,z}\left(z_{i_3}-z_{i_1}\right)}}{(p-q_3)^2+i\e}\,J\left(E,\tvec{p}-\tvec{q}_1-\tvec{q}_2-\tvec{q}_3\right) \,.
\end{align}
The integral appearing due to the contact interaction is given by
\begin{align}
\mathcal{I}^{21}_{i_1}&=2E\int\frac{dq_{2,z}}{(2\pi)}\frac{v_{i_1}\left(\tvec{q}_1,\frac{(p-q_1-q_2-q_3)_\perp^2-p_\perp^2}{2E}-q_{2,z}-q_{3,z}\right)v_{i_1}\left(\tvec{q}_2,q_{2,z}\right)}{\left(p-q_2-q_3\right)^2+i\e}\notag\\
&\simeq-\frac{i}{2}v_{i_1}\left(q_{1\perp}^2\right)v_{i_1}\left(q_{2\perp}^2+q_{3,z}^2\right)+\frac{\tilde{\m}_1+2\tilde{\m}_2}{\tilde{\m}_1\tilde{\m}_2(\tilde{\m}_2^2+q_{3,z}^2)\left((\tilde{\m}_1+\tilde{\m}_2)^2+q_{3,z}^2\right)}q_{3,z}\,,
\end{align}
exhibiting a non-zero real part even at the leading order in the eikonal expansion. Here $\tilde{\m}_n^2\equiv\m^2+\tvec{q}_n^2$, and we keep the subscript $i_1$ implicit in $\tilde{\m}$. One can notice that $\mathcal{I}^{21}_{i_1}$ results only in $q_{3,z}$-poles, leading to exponentially suppressed contributions, and we can easily perform the last $q_z$-integration. Then,
\begin{align}
iM_{21} &=\mathcal{C}\sum_{i_1,i_3}\int \frac{d^2 \tvec{q}_1\,d^2 \tvec{q}_2\,d^2 \tvec{q}_3}{(2\pi)^6}\, \left(t^a_\mathrm{proj} t_{i_3}^a\right)\,\theta_{2,1}\theta_{1,0}\,e^{-i (\tvec{q}_{1}+\tvec{q}_{2}) \cdot \tvec{x}_{i_1}}e^{-i \tvec{q}_{3} \cdot \tvec{x}_{i_3}}  \mathcal{I}^{12}_{i_1}\left(\tvec{q}_1,\tvec{q}_2\right) 
    \notag\\
    &\hspace{1cm}\times e^{-i\frac{(p-q_1-q_2-q_3)_\perp^2-p_\perp^2}{2E}z_{i_1}}e^{-i\frac{(p-q_3)_\perp^2-p_\perp^2}{2E}\left(z_{i_3}-z_{i_1}\right)}\,J\left(E,\tvec{p}-\tvec{q}_1-\tvec{q}_2-\tvec{q}_3\right) \,,
\end{align}
where we have used that $\text{Re} \, \mathcal{I}^{21}$ is sub-eikonal after the $q_{3,z}$ is substituted (the two integrals $\mathcal{I}^{12}$ and $\mathcal{I}^{21}$ coincide in this limit). Turning to the contribution to the squared amplitude, we find
\begin{align}
\left\langle M_{21}M^*_1 \right\rangle + c.c. &=-\mathcal{C}^2\int\limits^L_0\,dz_2\int\limits^{z_2}_0\,dz_1\,\int \frac{d^2 \tvec{q}_1\,d^2 \tvec{q}_2}{(2\pi)^4} \Bigg\{1+\frac{\tvec{p}-\tvec{q}_2}{E}z_1\cdot \hat{\tvec{g}}_1+\frac{\tvec{p}-\tvec{q}_2}{E}z_1\cdot \hat{\tvec{g}}_2\notag\\
&\hspace{0cm}+\frac{\tvec{p}-\tvec{q}_2}{E}(z_2-z_1)\cdot\hat{\tvec{g}}_2\Bigg\}\,\r_2\,\r_1\,v_2^2(q_{2\perp}^2)v_1^2(q_{1\perp}^2)|J(E, \tvec{p}-\tvec{q}_2)|^2\,,
\end{align}
where we have adjusted the integration variables to match the form of the other contributions.

Finally, we have to consider two types of diagrams involving two DB interactions. The first of such amplitude reads
\begin{align}
\left\langle iM_{22} \right\rangle &=\frac{1}{4}\mathcal{C}^2\sum_{i_1,i_3}\int \frac{d^2 \tvec{q}_1\,d^2 \tvec{q}_2\,d^2 \tvec{q}_3\,d^2 \tvec{q}_4}{(2\pi)^8}\,\theta_{2,1}\theta_{1,0}\,e^{-i \left(\tvec{q}_{1}+\tvec{q}_{2}\right) \cdot \tvec{x}_{i_1}}e^{-i \left(\tvec{q}_{3}+\tvec{q}_{4}\right) \cdot \tvec{x}_{i_3}}
    \notag\\
    &\hspace{0.5cm}\times  \, e^{-i\frac{(p-q_1-q_2-q_3-q_4)_\perp^2-p_\perp^2}{2E}z_{i_1}}e^{-i\frac{(p-q_3-q_4)_\perp^2-p_\perp^2}{2E}\left(z_{i_3}-z_{i_1}\right)}\notag\\
    &\hspace{1cm}\times v_{i_1} (q_{1\perp}^2)v_{i_1} (q_{2\perp}^2)v_{i_3} (q_{3\perp}^2)v_{i_3} (q_{4\perp}^2)\,J\left(E,\tvec{p}-\tvec{q}_1-\tvec{q}_2-\tvec{q}_3-\tvec{q}_4\right) \,,
\end{align}
where we have used the properties of the contact $q_z$-integrals from $M_{12}$ and $M_{21}$ omitting unnecessary details. Combining with the conjugated amplitude, we write
\begin{align}
\left\langle M_{22}M^*_0 \right\rangle + c.c. &=\frac{1}{2}\mathcal{C}^2\int\limits^L_0\,dz_2\int\limits^{z_2}_0\,dz_1\,\int \frac{d^2 \tvec{q}_1\,d^2 \tvec{q}_2}{(2\pi)^4} \Bigg\{1+\frac{\tvec{p}}{E}z_1\cdot \hat{\tvec{g}}_1+\frac{\tvec{p}}{E}z_1\cdot \hat{\tvec{g}}_2\notag\\
&\hspace{1cm}+\frac{\tvec{p}}{E}(z_2-z_1)\cdot\hat{\tvec{g}}_2\Bigg\}\,\r_2\,\r_1\,v_2^2(q_{2\perp}^2)v_1^2(q_{1\perp}^2)|J(E, \tvec{p})|^2\,,
\end{align}
where the numeration of momenta has been changed for convenience. 

Similarly, one has to take into account the DB amplitude squared. Using the result of \cite{Sadofyev:2021ohn} we write
\begin{align}
\left\langle M_{2}M^*_2 \right\rangle &=\frac{1}{4}\mathcal{C}^2\int\limits^L_0\,dz_2\int\limits^{L}_0\,dz_1\,\int \frac{d^2 \tvec{q}_1\,d^2 \tvec{q}_2}{(2\pi)^4} \Bigg\{1+\frac{\tvec{p}}{E}z_1\cdot \hat{\tvec{g}}_1+\frac{\tvec{p}}{E}z_2\cdot \hat{\tvec{g}}_2\Bigg\}\notag\\
&\hspace{1cm}\times\r_2\,\r_1\,v_2^2(q_{2\perp}^2)v_1^2(q_{1\perp}^2)|J(E, \tvec{p})|^2\,,
\end{align}
and notice that the $z$-integration limits differ from the limits in other diagrams. In fact, in the case of a longitudinally invariant matter, it can be easily seen that this pair of integrals is twice larger than the ordered pair of $z$-integrals. 

Thus, collecting all the contributions, we find the full amplitude squared at $N=2$, and it reads
\begin{align}
\left\langle \left| M\right|^2 \right\rangle^{(2)} & = \int\displaylimits_0^{L}\, dz_2\int\displaylimits_0^{z_2}\, dz_1\int \frac{d^2\tvec{q}}{(2\pi)^2}
 \Bigg\{1+\frac{\tvec{p}-\tvec{q}_1-\tvec{q}_2}{E}z_1\cdot \hat{\tvec{g}}_1+\frac{\tvec{p}-\tvec{q}_1-\tvec{q}_2}{E}z_1\cdot \hat{\tvec{g}}_2\notag\\
&+\frac{\tvec{p}-\tvec{q}_2}{E}(z_2-z_1)\cdot\hat{\tvec{g}}_2\Bigg\}\mathcal{V}_1(\tvec{q}_1)\mathcal{V}_2(\tvec{q}_2)|J(E, \tvec{p}-\tvec{q}_1-\tvec{q}_2)|^2\,,
\end{align}
in agreement with the general expression \eqref{MR1}. Repeating the same calculation for the higher orders in opacity, we find that the structure of \eqref{MR1} is reproduced. 

To understand the origin of the all order pattern, one may consider the different classes of the contributions to the amplitude squared, distinguished by the color pairing. For the given pairing, one can consider three higher order diagrams obtained by attaching two extra in-medium field insertions in all possible ways -- one extra SB interaction on each of the sides of the cut and one extra DB interaction on either one of the sides of the cut. Thus, for each color pairing at $N=n$ we obtain a set of contributions at $N=n+1$. Comparing to the $N=2$ case, we can see that each of these contributions has the same structure as in the case of the already studied diagrams. Paying a particular attention to the LPM phases, we can see that at the first order in gradients, all the phases may depend only on the momenta of the two additional interactions which are responsible for the gradient contributions arising at $N=n+1$. 

\section{Shift Operator and $\d$-function}
\label{App2}

In this Appendix, we briefly discuss the imaginary term in the EOMs \eqref{eq:ode}, arising due to the gradient effects. This term results in an imaginary shift of the zeroth order trajectory, which enters for instance into the argument of $\d$-functions in the last line of \eqref{eq:GG_some}. Since the $\d$-functions are originally defined in the real space, this notation is only formal, and the functions should be re-defined.

In the main text, the resulting delta functions are treated as in the case of a real argument. To justify this treatment, consider the following one-dimensional integral 
\begin{align}
\int dx \, e^{-ipx} e^{ a x} \equiv (2\pi)\, \delta(p+i a)  \, .
\end{align}
For a real $a$, one can understand it as a momentum representation of the shift operator $e^{a x}=e^{ia \frac{\pa}{\pa p}}$ with a purely imaginary argument $ia$. Thus, we can write
\begin{align}
   \int dx \, e^{-ipx}\,  e^{a x  } = (2\pi)e^{ia \frac{\pa}{\pa p}}\d(p)=(2\pi)\, \delta(p+i a) \,  \,.
\end{align}
This simple calculation shows that the imaginary term entering the EOM will indeed give rise to a momentum-space shift operator, and its effect can be formally written with the complex argument of a $\d$-delta. For instance, the one-dimensional analogue to the integral of interest in the main text reduces to
\begin{align}
   &\int dp_{in }d\bar p_{in } \,  \delta(\bar p_{in }-p_{in }-ia) J(p_{in })J^\dagger(\bar p_{in })  \nn 
   &=\int dp_{in } \,  |J(p_{in})|^2 + ia  \int dp_{in }d\bar p_{in } \,  (2\pi)\, \delta(\bar p_{in}-p_{in})\frac{\partial }{\partial(\bar p_{in}-p_{in})} \left[ J(p_{in})J^\dagger(\bar p_{in} ) \right] \nn 
   &= \int dp_{in } \,  |J(p_{in})|^2  + \frac{i a}{2}  \int dq dQ \, (2\pi)\, \delta(q) \frac{\partial}{\partial q} \left[ J\left(\frac{Q-q}{2}\right)J^\dagger\left(\frac{Q+q}{2}\right) \right]\nn 
   &=\int dp_{in } \,  |J(p_{in})|^2  \, .
\end{align}

\bibliographystyle{bibstyle}
\bibliography{references}

\end{document}